\documentclass{aa}

\usepackage{txfonts}
\usepackage{graphicx}

\begin{document}

\title{Phase Mixing of Nonlinear Visco-resistive Alfv\'en Waves}

\author{J.~A.~McLaughlin\thanks{Present address: School of Computing and Engineering, Northumbria University, Newcastle Upon Tyne, NE1 8ST}, I.~De Moortel \and A.~W. Hood}

\offprints{J.~A.~McLaughlin, \email{james.a.mclaughlin@northumbria.ac.uk}}

\institute{{School of Mathematics and Statistics, University of St Andrews, KY16 9SS, UK}}

\date{Received 9 August 2010 / Accepted 11 January 2011}

\authorrunning{McLaughlin {{et al.}}}
\titlerunning{Nonlinear Alfv\'en Wave Propagation and Phase Mixing}

%----------------------------------------------------------------

\abstract
{}
{We investigate the behaviour of nonlinear, nonideal Alfv\'en wave propagation within an inhomogeneous magnetic environment.}
{The governing MHD equations are solved in 1D and 2D using both analytical techniques and numerical simulations.}
{We find clear evidence for the ponderomotive effect and visco-resistive heating. The ponderomotive effect generates a longitudinal component to the transverse Alfv\'en wave, with a frequency twice that of the driving frequency. Analytical work shows the addition of resistive heating. This leads to a substantial increase in the local temperature and thus gas pressure of the plasma, resulting in material being pushed along the magnetic field. In 2D, our system exhibits phase mixing and we observe an evolution in the location of the maximum heating, i.e. we find a  drifting of the heating layer.}
{Considering Alfv\'en wave propagation in 2D with an inhomogeneous density gradient, we find that the equilibrium density profile is significantly modified by {\emph{both}} the flow of density due to visco-resistive heating {\emph{and}} the nonlinear response to the localised heating through phase mixing.}

\maketitle

%----------------------------------------------------------------

\section{Introduction}\label{section1}

Phase mixing, a mechanism for dissipating shear Alfv\'en waves, was first proposed by  Heyvaerts \& Priest (\cite{Heyvaerts1983}). The basic concept is straightforward: consider shear Alfv\'en waves propagating in an inhomogeneous plasma, such that  on each magnetic fieldline each wave propagates with its own local Alfv\'en speed. Thus, after propagating a certain distance, these neighbouring perturbations will be out of phase, which will lead to the generation of smaller and smaller transverse spatial scales, and thus to the growth of strong currents. This ultimately results in a strong enhancement of the dissipation of Alfv\'en-wave energy via viscosity and/or resistivity. Alfv\'en wave phase mixing has also been studied extensively as a possible mechanism for heating the corona (e.g. Heyvaerts \& Priest \cite{Heyvaerts1983}, Browning \cite{Browning1991}, Ireland \cite{Ireland1996}, Malara {{et al.}} \cite{Malara1996}, Narain \& Ulmschneider  \cite{Narain1990}; \cite{Narain1996}).

Nakariakov {{et al.}} (\cite{Nakariakov1997}) extended the model of  Heyvaerts \& Priest (\cite{Heyvaerts1983}) to include compressibility and nonlinear effects. They found that fast magnetoacoustic waves are generated continuously by Alfv\'en-wave phase mixing at a frequency twice that of the driven Alfv\'en wave, and that these generated waves propagate across magnetic fieldlines and away from the phase mixing layer. Since there is a permanent leakage of energy away from the phase mixing layer,  these fast waves can cause indirect heating of the plasma as they propagate away and dissipate far from the layer itself, thus spreading the heating across the domain. Nakariakov {{et al.}} (\cite{Nakariakov1997}) found that the amplitude of these fast waves grows linearly in time, according to weakly-nonlinear, $\beta=0$, analytical theory. Nakariakov {{et al.}} (\cite{Nakariakov1998}) further extended this model to include a background  steady flow and found that the findings of their  \cite{Nakariakov1997} model persist.

Hood {{et al.}} (\cite{Hood2002}) investigated the phase mixing of single Alfv\'en-wave pulses and found that this results in a slower power-law damping (as opposed to the standard $\sim \exp{(-t^3)}$ for harmonic Alfv\'en waves, i.e. Heyvaerts \& Priest \cite{Heyvaerts1983}). Hence,  Alfv\'enic-pulse perturbations will be able to transport energy   to a greater coronal height  than that of a harmonic Alfv\'en wavetrain.

Botha {{et al.}} (\cite{Botha2000})  considered a developed stage of Alfv\'en-wave phase mixing and found that the growth of the generated fast waves saturates at amplitudes much lower than that of the driven Alfv\'en wave. They concluded that the nonlinear generation of fast waves (Nakariakov {{et al.}} \cite{Nakariakov1997}) {{saturates}} due to destructive wave interference, and has little effect on the standard phase mixing model of  Heyvaerts \& Priest (\cite{Heyvaerts1983}).  The numerical simulations of Botha {{et al.}} assumed wave propagation in an ideal plasma, and thus heating was absent from their model. Tsiklauri and co-authors repeated these numerical experiments for both a {{weakly}} and strongly-nonlinear Alfv\'enic pulse (Tsiklauri {{et al.}} \cite{Tsiklauri2001}; \cite{Tsiklauri2002}).

De Moortel {{et al.}} (\cite{DeMoortel1999}; \cite{DeMoortel2000}) investigated how gravitational density stratification and magnetic field divergence changes the efficiency of phase mixing. They report that the resultant dissipation can be either enhanced or diminished depending on the specific choice of equilibrium. Ruderman {{et al.}} (\cite{Ruderman1998}) investigated phase mixing in open magnetic equilibria, and found similar results using the WKB method. These \cite{Ruderman1998} analytical results were repeated and corrected by numerical and semi-analytical work by Smith {{et al.}} (\cite{Smith2007}). All these authors found the following: a diverging magnetic field enhances the efficiency of phase mixing, whereas gravitational stratification diminishes the mechanism. Hood {{et al.}} (\cite{Hood1997a}; \cite{Hood1997b}) derive analytical, self-similar solutions of Alfv\'en-wave phase mixing in both open and closed magnetic topologies.

%%%%%%%%%%%%%%%%%%%%%%%%%%%%%%%%%%%%%%%%%%%%%%%%%%%%%%%%%%%%%%%%%%%%%%%%%%%%%%%

Ofman \& Davila (\cite{Ofman1995}) found that in an inhomogeneous coronal hole with an enhanced dissipation parameter $(S=1000-10,000)$, Alfv\'en waves can dissipate within several solar radii, which can provide significant energy for the heating and acceleration of the solar wind. This model was later extended to include nonlinear effects (Ofman \& Davila \cite{Ofman1997}).

Parker (\cite{Parker1991}) pointed out that phase mixing requires an ignorable coordinate, an assumption which is expected to be unphysical in the corona. Parker found that including all three coordinates results in the driven Alfv\'en waves coupling with a fast magnetoacoustic mode, and that this elimates the growth in transverse spatial scales, and thus phase mixing is absent from the system. Instead, such a system exhibits resonant absorption (e.g. Lee \& Roberts \cite{Lee1986}; Hollweg \& Yang \cite{Hollweg1988}) on the surfaces where the phase velocity equals the Alfv\'en velocity. However, Parker's conclusions are in disagreement with the work of Tsiklauri and co-authors who considered the interaction of an impulsively-generated, weakly-nonlinear MHD pulse with a one-dimensional density inhomogeniety, considered in the three-dimensional regime (i.e. without an ignorable coordinate) in both an ideal (Tsiklauri \& Nakariakov \cite{TN2002}) and resistive (Tsiklauri {{et al.}} \cite{Tsiklauri2003}) plasma. Tsiklauri and co-authors  found that phase mixing remains a relevant paradigm and that the dynamics can still be qualitatively understood in terms of the classic 2.5D models. Mocanu {{et al.}} (\cite{Mocanu2008}) have revisited  the Heyvaerts \& Priest (\cite{Heyvaerts1983}) model using anisotopic viscosity (i.e. incorporating the Braginskii \cite{Braginskii1965} stress tensor) and report that this significantly increases the damping lengths, i.e. compared to those obtained for isotropic dissipation. More recently, Threlfall {{et al.}} (\cite{Threlfall2010}) have investigated the effect of the Hall term on phase mixing in the ion-cyclotron range of frequencies.

%%%%%%%%%%%%%%%%%%%%%%%%%%%%%%%%%%%%%%%%%%%%%%%%%%%%%%%%%%%%%%%%%%%%%%%%%%%%%%%

Another key concept that we shall invoke in this paper is the ponderomotive force: a nonlinear force proportional to spatial gradients in magnetic pressure, also referred to as the Alfv\'en wave-pressure force. The ponderomotive effect  has been considered in a solar context initially by Hollweg (\cite{Hollweg1971}) and later by Verwichte and co-authors (Verwichte \cite{VerwichteTHESIS1999}; Verwichte {{et al.}} \cite{VNL1999}). Hollweg (\cite{Hollweg1971}) {{considered}}  linearly-polarised Alfv\'en waves propagating in a direction parallel to the magnetic field, and found that the transverse behaviour of the Alfv\'en wave was identical when comparing the linear and nonlinear (to second order) solutions, but that  longitudinal wave velocity and density fluctuations appear in the nonlinear solutions  driven by gradients in the wave magnetic-field pressure, i.e. the Alfv\'en wave is no longer purely transverse and is compressive (through nonlinear coupling to magnetoacoustic waves). Thus, the ponderomotive force can be used as an extra acceleration mechanism and as an explanation for density fluctuations in the solar wind. Verwichte (\cite{VerwichteTHESIS1999}) presented a mechanical analogy for the ponderomotive effect by considering the resulting motion of discrete particles (beads)  on an oscillating string. Verwichte {{et al.}} (\cite{VNL1999}) considered the temporal evolution of  a weakly-nonlinear, Alfv\'en wave in a $\beta=0$ homogeneous plasma. These authors showed that the an initially-excited, gaussian-pulse perturbation in transverse velocity splits into two Alfv\'en wave pulses, each propagating in opposite directions (as naturally expected). Furthermore,  Verwichte {{et al.}} found that the ponderomotive force produces a shock in longitudinal velocity at the starting position. Note that in a cold plasma, there is no force to counteract this ponderomotive acceleration.

In this paper, we investigate the  nonlinear, nonideal behaviour of Alfv\'en-wave propagation and phase mixing over long timescales. Thus, this paper can be seen as an extension of the model of Botha {{et al.}} (\cite{Botha2000})  to include  visco-resistive effects. We also seek to address a fundamental question of phase mixing:  by considering nonlinear, nonideal phase mixing over long timescales, {\emph{is it possible to observe a drifting of the heating layer?}} In other words,  phase mixing will occur due to the density inhomogenity in our system, and this process will generate strong, localised heating due to enhanced dissipation. This localised heat deposition is expected to modify the equilibrium density profile, and thus may change the location of maximum heating. However, it is unclear what the actual result will be: we may observe a change in the location of the heating layer, or the phase mixing mechanism may break down, or the heating may bifurcate spatially (since our density profile will no longer be monotonic). Is is also unclear how this may affect the indirect heating of the plasma, due to the coupling to the fast magnetoacoustic mode (Nakariakov {{et al.}} \cite{Nakariakov1997}). Such an investigation requires nonlinear and nonideal effects to be considered and, as we shall see,  observed  over long timescales.

% Thus, this paper is the first such investigation to consider these three factors together.

The work of Ofman {{et al.}} (\cite{Ofman1998}) is also relevant here. These authors investigated a model of resonant absorption that incorporated the dependence of loop density on the heating rate, and studied the spatial and temporal dependence of the heating layer. Ofman {{et al.}} find that the heating occurs in multiple resonance layers, rather than the single layer of the classic resonant absorption models (e.g. Ionson \cite{Ionson1978}; Ulmschneider {{et al.}} \cite{Ulmschneider1991}; Ruderman \& Roberts \cite{RR2002}) and that these layers drift throughout the loop to heat the entire volume. Poedts \& Boynton (\cite{Poedts1996}) also investigated resonant absorption using nonlinear, resistive MHD simulations and found a spreading of the heat deposition, i.e. a broadening of the resonant layer due to changes in the background inhomogeneity.

%In a numerical experiment in non-linear resistive MHD, Poedts & Boynton (1996) excited the AlfvÂ´en waves directly at the footpoints and found that nonlinear AlfvÂ´en waves can dissipate sufficient energy to heat coronal loops without resorting to the resonant coupling to a quasi mode.

This paper has the following outline: $\S\ref{section2}$ describes the governing equations, assumptions and analytical and numerical details of our investigation, $\S\ref{section:1-D}$ investigates the 1D nonlinear, nonideal system (with no density inhomogeneity) and focuses on the underlying physical processes.  $\S\ref{section:bulk_flow}$ details the bulk-flow phenomenon found to be present in our system, and investigates its density dependence. $\S\ref{section:2-D}$ considers a 2D model with  an inhomogeneous density profile, and details the long-term evolution and coupled nature of the three MHD waves present in our system. The conclusions are presented in $\S\ref{section:conclusions}$ and there are two appendicies.

%----------------------------------------------------------------

\section{Basic Equations}\label{section2}

We consider the nonlinear, compressible, viscous and resistive MHD equations appropriate to the solar corona:
\begin{eqnarray}
{\partial \rho\over \partial t} + \nabla \cdot \left (\rho {\bf{v}}\right ) &=& 0\; ,\nonumber  \\
\rho \left[ {\partial {\bf{v}}\over \partial t} + \left( {\bf{v}}\cdot\nabla \right) {\bf{v}} \right] &=& - \nabla p + \left( {{\frac{   \nabla \times {\bf{B}}    }{\mu}}}    \right)\times {\bf{B}}   + \nu \nabla \cdot {\bf{S}}   \; ,\nonumber \\
 {\partial {\bf{B}}\over \partial t}  &=& \nabla \times \left ({\bf{v}}\times {\bf{B}}\right ) + \eta \nabla ^2  {\bf{B}}\; , \label{MHDequations}   \\
{\partial {p}\over \partial t}  + \left( {\bf{v}}\cdot\nabla \right) p &=& - \gamma p \nabla \cdot {\bf{v}} + {{\frac{\gamma -1}{\sigma}}} \left| {\bf{j}} \right| ^2 + \nu \left(\gamma -1 \right)  Q_{visc}  \;,    \nonumber 
\end{eqnarray}
where $\rho$ is the mass density, ${\bf{v}}$ is the plasma velocity, ${\bf{B}}$ the magnetic induction (usually called the magnetic field), $p$ is the plasma pressure,  $ \mu = 4 \pi \times 10^{-7} \/\mathrm{Hm^{-1}}$  is the magnetic permeability, $\sigma$ is the electrical conductivity,  $\eta=1/ {\mu \sigma} $ is the magnetic diffusivity, $\gamma={5 / 3}$ is the ratio of specific heats and ${\bf{j}} = {{\nabla \times {\bf{B}}} / \mu}$ is the electric current density. In this paper, $\eta$ and $\nu$ are assumed to be constants.

The viscous stress tensor, ${\bf{S}}$, is given by
\begin{eqnarray*}
S_{ij} = \epsilon_{ij} -\frac{1}{3}\delta_{ij} \nabla \cdot {\bf{v}}  \;, \nonumber %\label{eq:visc_tensor}
\end{eqnarray*}
where ${\bf{\epsilon}}$ is the rate-of-strain tensor, given by:
\begin{eqnarray*}
\epsilon_{ij} = \frac{1}{2}\left (\frac{\partial v_i}{\partial x_j} + \frac{\partial v_j}{\partial x_i}\right ) \;.\nonumber %\label{eq:strain}
\end{eqnarray*}

The viscous heating term is given by
\begin{eqnarray*}
Q_{visc} = \sum_{i=1}^3 \sum_{j=1}^3 \epsilon_{ij}S_{ij}\;. \nonumber %\label{eq:viscous_heating}
\end{eqnarray*}

Note that in the presence of strong magnetic fields,  the classical viscous term used in equations (\ref{MHDequations}) is not the most appropriate for the solar corona, since the viscosity takes the form of a non-isotropic tensor. The mathematical details of this non-isotropic tensor can be found in Braginskii (\cite{Braginskii1965}). {{In this paper, we have chosen to implement the simplest, isotropic version of the stress tensor as a first step in investigating nonlinear phase mixing, and non-isotropic viscosity will be considered in future investigations. Our choice of viscosity term also allows a  direct comparison with the results of  Heyvaerts \& Priest (\cite{Heyvaerts1983}) and allows us to derive analytical solutions to our governing equations.}}

We now consider a change of scale to non-dimensionalise all variables. Let ${\rm{\bf{v}}} = {\rm{v}}_0 {\mathbf{v}}^*$,  ${\mathbf{B}} = B {\mathbf{B}}^*$, $x = L x^*$, $y=L y^*$, $\rho={\rho}_0 \rho^*$, $p = p_0 p^*$, $\nabla = \frac{1}{L}\nabla^*$, $t={\tau}_A t^*$, $\eta = \eta_0$ and  $\nu = \nu_0$, where * denotes a dimensionless quantity and ${\rm{v}}_0$, $B$, $L$, ${\rho}_0$, $p_0$, ${\tau}_A$, $\eta_0$ and $\nu_0$ are constants with the dimensions of the variable they are scaling. We then set $ {B} / {\sqrt{\mu \rho _0 } } ={\rm{v}}_0$ and ${\rm{v}}_0 =  {L} / {{\tau_A}}$ (this sets ${\rm{v}}_0$ as a constant equilibrium Alfv\'{e}n speed). We also set ${\eta_0 {\tau}_A } /  {L^2} =R_m^{-1}$, where $R_m$ is the magnetic Reynolds number. This process non-dimensionalises equations (\ref{MHDequations}) and under these scalings, $t^*=1$ (for example) refers to $t={\tau}_A=  {L} / {{\rm{v}}_0}$; i.e. the time taken to travel a distance $L$ at the background Alfv\'en speed. For the rest of this paper, we drop the star indices; the fact that all variables are now non-dimensionalised is understood.

%and set $ {\beta_0} = {2 \mu p_0} / {B^2}$, where $\beta_0$ is the plasma-$\beta$ at a radius $L$ from the origin.

\subsection{Basic equilibrium}\label{section:2.1}

We consider equations (\ref{MHDequations}) in Cartesian coordinates and assume there are no variations in the $z-$direction (${{\partial} /{\partial z}} =0$). We consider an inhomogeneous plasma in the $x-$direction embedded in a uniform magnetic field in the $y-$direction, i.e.:
\begin{eqnarray}
\rho_0 = \rho_0(x)\;,\; p_0= {\rm{constant}}\;,\;{\bf{B}}_0 = (0,B_0,0) \;,\; {\bf{v}}_0 = {\bf{0}}\label{B_0}
\end{eqnarray}
with finite amplitude perturbations of the form:
\begin{eqnarray}
\rho&=&\rho(x,y,t) \;,\; {\bf{v}}=({{\rm{v}}_x},{{\rm{v}}_y},{{\rm{v}}_z})= {\bf{v}}(x,y,t)\nonumber\\
p&=&p(x,y,t) \;,\; {\bf{B}}=(B_x,B_y,B_z)= {\bf{B}}(x,y,t)\label{pert2}
\end{eqnarray}

\begin{figure*}
\begin{center}
\includegraphics[width=7.2in]{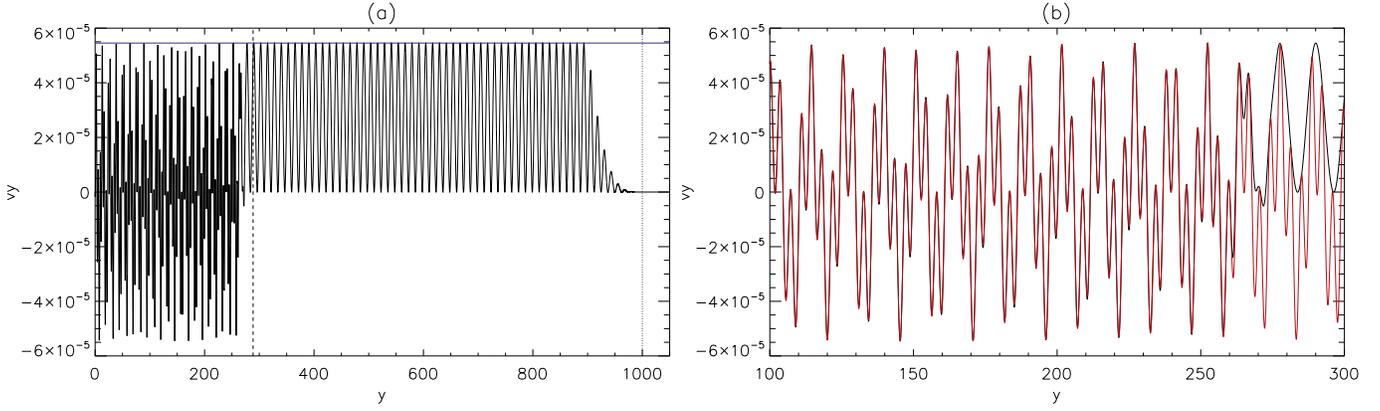}
\caption{$(a)$  Longitudinal perturbations ${{\rm{v}}_y}$ for  $\beta=0.1$ plasma with $\eta=\nu=0$. Dotted and dashed lines represent $y={\rm{v}}_At$ and $y=c_st$ respectively. The blue line denotes the maximum amplitude: $y=A^2/4 \left( {\rm{v}}_A^2- c_s^2 \right)$. $(b)$ Blow-up of region $100\le y \le 200$ for ${{\rm{v}}_y}$. In both subfigures, $t=1000\tau_A$.}
\label{figure2}
\end{center}
\end{figure*}

\subsection{Analytical description}\label{section:2.2}
Following the analysis of Nakariakov {{et al.}} (\cite{Nakariakov1995}; \cite{Nakariakov1997}) and Botha {{et al.}} (\cite{Botha2000}), equations (\ref{MHDequations}) are written in component form using equilibrium (\ref{B_0}) and perturbations (\ref{pert2}):
\begin{eqnarray}
{{\partial \rho}\over{\partial t}} + {{\partial}\over{\partial x}} \left( \rho_0 {{\rm{v}}_x}\right) + \rho_0 {{\partial {{\rm{v}}_y}}\over {\partial y}} &=& N_1 \label{rho}\;,\\
\rho_0 {{\partial {{\rm{v}}_x}}\over {\partial t}} + {{\partial p}\over {\partial x}} + {{B_0} \over{ \mu}} \left( {{\partial B_y }\over {\partial x}} - {{\partial B_x}\over {\partial y}} \right) + V_1&=& N_2 \label{v_x}\;,\\
\rho_0 {{\partial {{\rm{v}}_y}}\over {\partial t}} + {{\partial p}\over {\partial y}} + V_2&=& N_3 \label{v_y}\;,\\
\rho_0 {{\partial {{\rm{v}}_z}}\over {\partial t}} - {{B_0}\over{\mu}} {{\partial B_z }\over {\partial y}} +V_3 &=& N_4 \label{v_z}\;,\\
{{\partial B_x }\over {\partial t}} - {B_0} {{\partial {{\rm{v}}_x}}\over {\partial y}} - R_1 &=& N_5 \label{B_x}\ \;,\\
{{\partial B_y}\over {\partial t}} + B_0 {{\partial {{\rm{v}}_x} }\over {\partial x}} - R_2 &=& N_6 \label{B_y} \;,\\
{{\partial B_z}\over {\partial t}} - B_0  {{\partial {{\rm{v}}_z}}\over {\partial y}} - R_3 &=& N_7 \label{B_z} \;,\\ 
{{\partial p }\over {\partial t}} + \gamma p_0 \left( {{\partial {{\rm{v}}_x}}\over {\partial x}}+ {{\partial {{\rm{v}}_y} }\over {\partial y}}\right)  &=& N_8 \label{P}\;,
\end{eqnarray}
where the left-hand-sides involve only linear terms and $N_1$-$N_8$ denote  the nonlinear terms. The nonlinear terms are:
\begin{eqnarray}
N_1 &=& - {{\partial }\over {\partial x}} \left( \rho {{\rm{v}}_x} \right) - {{\partial }\over {\partial y}} \left( \rho {{\rm{v}}_y} \right) \;,\\
N_2 &=& -  {{B_z}\over{\mu}} {{\partial B_z}\over {\partial x}} - \rho {{\partial {{\rm{v}}_x}}\over {\partial t}} - \left( \rho_0 + \rho\right) \left( {{\rm{v}}_x} {{\partial }\over {\partial x}} + {{\rm{v}}_y} {{\partial }\over {\partial y}} \right) {{\rm{v}}_x} \nonumber \;\\
&&\quad-{{B_y} \over{\mu}} \left( {{\partial B_y}\over {\partial x}} - {{\partial B_x}\over {\partial y}} \right) \;,\\
N_3 &=& - {{B_z}\over{\mu}}{{\partial B_z}\over {\partial y}}-\rho {{\partial {{\rm{v}}_y}}\over {\partial t}} - \left(\rho_0 + \rho\right) \left( {{\rm{v}}_x} {{\partial }\over {\partial x}} + {{\rm{v}}_y} {{\partial }\over {\partial y}} \right) {{\rm{v}}_y}  \nonumber \;\\
&&\quad+ {{B_x} \over{\mu}} \left( {{\partial B_y }\over {\partial x}} - {{\partial B_x}\over {\partial y}} \right) \;, \\
N_4 &=& - \rho {{\partial {{\rm{v}}_z} }\over {\partial t}} - \left( \rho_0 + \rho \right) \left( {{\rm{v}}_x} {{\partial }\over {\partial x}} + {{\rm{v}}_y} {{\partial }\over {\partial y}} \right) {{\rm{v}}_z} \nonumber \\
&&\quad+ {{B_y}\over{\mu}} {{\partial B_z}\over {\partial y}} + {{B_x} \over{\mu}} {{\partial B_z}\over {\partial x}}\;, \\
N_5 &=& {{\partial }\over {\partial y}} \left( {{\rm{v}}_x} B_y - {{\rm{v}}_y} B_x \right)\;,\\
N_6 &=& - {{\partial }\over {\partial x}}  \left( {{\rm{v}}_x}B_y - {{\rm{v}}_y}B_x \right)\;,\\
N_7 &=& {{\partial }\over {\partial x}}  \left( {{\rm{v}}_z}B_x - {{\rm{v}}_x} B_z \right) + {{\partial }\over {\partial y}}  \left( {{\rm{v}}_z}B_y - {{\rm{v}}_y} B_z \right)\;,\\
N_8 &=& -\left( {{\rm{v}}_x} {{\partial }\over {\partial x}} + {{\rm{v}}_y} {{\partial }\over {\partial y}} \right) p - \gamma p \left( {{\partial {{\rm{v}}_x}}\over {\partial x}} + {{\partial {{\rm{v}}_y}}\over {\partial y}} \right) + R_4 + V_4\;,
\end{eqnarray}
the resistive terms (where $R_1$-$R_3$ are linear, $R_4$ is nonlinear) are:
\begin{eqnarray}
R_1 &=& \eta \left( {{\partial^2 }\over {\partial x^2}} + {{\partial ^2}\over {\partial y^2}} \right) B_x \label{R1} \;,\\
R_2 &=& \eta \left( {{\partial^2 }\over {\partial x^2}} + {{\partial ^2}\over {\partial y^2}} \right) B_y \label{R2} \;,\\
R_3 &=& \eta \left( {{\partial^2 }\over {\partial x^2}} + {{\partial ^2}\over {\partial y^2}} \right) B_z \label{R3} \;,\\
R_4  &=& \left( \gamma -1 \right) {{{\eta} \over {\mu}}}  \left[ \left( {{\partial B_z}\over {\partial y}} \right) ^2 + \left( {{\partial B_y }\over {\partial x}} - {{\partial B_x}\over {\partial y}} \right) ^2 + \left( {{\partial B_z}\over {\partial x}} \right)^2 \right] \label{R4} \;,
\end{eqnarray}
and the viscous terms ($V_1$-$V_3$ are linear, $V_4$ is nonlinear)  are
\begin{eqnarray}
V_1 &=&-  \nu \left (  \frac{2}{3} \frac{\partial^2{{\rm{v}}_x}}{\partial x^2} +\frac{\partial^2 {{\rm{v}}_x}}{\partial y^2}    - \frac{1}{3}\frac{\partial^2 {{\rm{v}}_y}}{\partial x \partial y}                     \right ) \label{V1}\;,\\
V_2 &=& -  \nu \left (\frac{\partial^2{{\rm{v}}_y}}{\partial x^2} +   \frac{2}{3}     \frac{\partial^2 {{\rm{v}}_y}}{\partial y^2} - \frac{1}{3}\frac{\partial^2  {{\rm{v}}_x}}{\partial x \partial y}   \right )  \label{V2}\;,\\
V_3 &=& - \nu \left (\frac{\partial^2{{\rm{v}}_z}}{\partial x^2} +\frac{\partial^2 {{\rm{v}}_z}}{\partial y^2} \right )  \label{V3}\;,\\
V_4 &=& \nu \left( \gamma -1 \right)       \left\{   \left( {{\partial {{\rm{v}}_x}}\over {\partial x}} \right)^2  +   \frac{1}{2}  \left( {{\partial {{\rm{v}}_x} }\over {\partial y}} + {{\partial {{\rm{v}}_y}}\over {\partial x}} \right)^2     \right. \nonumber \\ 
&& \quad+   \frac{1}{2}  \left( {{\partial {{\rm{v}}_z}}\over {\partial x}} \right)^2       + \left.  \left( {{\partial {{\rm{v}}_y}}\over {\partial y}} \right)^2     +   \frac{1}{2}  \left( {{\partial {{\rm{v}}_z}}\over {\partial y}} \right)^2 \right. \nonumber \\
&& \quad  -\left.    \frac{1}{3} \left[  \left( {{\partial {{\rm{v}}_x}}\over {\partial x}} \right)^2 +   \left( {{\partial {{\rm{v}}_y}}\over {\partial y}} \right)^2  \right]  \left(   {{\partial {{\rm{v}}_x}}\over {\partial x}} + {{\partial {{\rm{v}}_y}}\over {\partial y}}     \right)     \right\}  \label{V4}\;.
\end{eqnarray}

\begin{figure*}
\begin{center}
\includegraphics[width=7.2in]{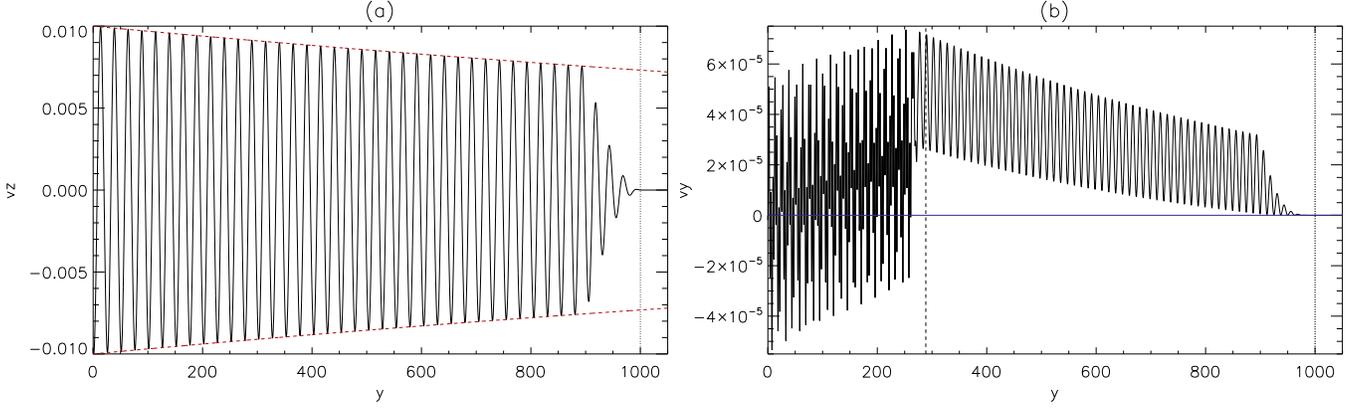}
\caption{$(a)$ Transverse perturbations ${{\rm{v}}_z}$ for $\eta=0.01$, $\beta=0.1$ plasma ($\nu=0$), where dashed lines represent the envelope ${{\rm{v}}_z}= \pm A \exp{\left(k_I y\right)}$. $(b)$ Longitudinal perturbations ${{\rm{v}}_y}$ for same plasma, where dashed line represents $y=c_s t$ and blue horizontal line denotes ${\rm{v}}_y=0$. In both subfigures, $t=1000\tau_A$ and  dotted line represents $y={\rm{v}}_At$.}
\label{figure3}
\end{center}
\end{figure*}

Equations (\ref{rho} - \ref{P}) describe all the waves occurring in the plasma. These equations can be combined to obtain the evolution expressions for the velocity components:
\begin{eqnarray}
&&\left[ {{\partial ^2}\over {\partial t^2}} - \left( {\rm{v}}_A^2 + c_S^2\right) {{\partial ^2 }\over {\partial x^2}} - {\rm{v}}_A^2 {{\partial ^2}\over {\partial y^2}} \right] {{\rm{v}}_x} - c_s^2 {{\partial ^2  {{\rm{v}}_y} }\over {\partial x \partial y}}  \nonumber \\
&&\qquad= {{1} \over{\rho_0}} \left[ - {\frac{\partial V_1}{\partial t}}  -{{\partial N_8}\over {\partial x}} + {{\partial N_2}\over {\partial t}} \right.\nonumber\\
&& \qquad- \left.{{B_0} \over{\mu}} \left( {{\partial N_6 }\over {\partial x}} + {{\partial R_2}\over {\partial x}}  -  {{\partial N_5  }\over {\partial y}} - {{\partial R_1}\over {\partial y}} \right) \right] \label{vxevolution}\;, \\
&&\left({{\partial^2} \over {\partial t^2}} - c_s^2 {{\partial ^2}\over {\partial y^2}} \right) {{\rm{v}}_y} - c_s^2 {{\partial ^2 {{\rm{v}}_x}}\over {\partial x \partial y}} \nonumber\\
&&\qquad= {{1} \over {\rho_0}} \left(  - \frac{\partial V_2}{\partial t}    +  {{\partial N_3}\over {\partial t}} - {{\partial N_8}\over {\partial y}} \right)\label{vyevolution}\;,\\
&&\left( {{\partial^2 }\over {\partial t^2}} - {\rm{v}}_A^2 {{\partial ^2}\over {\partial y^2}} \right) {{\rm{v}}_z} \nonumber \\
&&\qquad= {{1} \over{\rho_0}} \left[  - \frac{\partial V_3}{\partial t}   +     {{\partial N_4}\over {\partial t}} +   {{B_0}\over{\mu}} \left( {{\partial N_7 }\over {\partial y}} + {{\partial R_3}\over {\partial y}} \right) \right], \label{vzevolution}
\end{eqnarray}
where ${\rm{v}}_A^2(x)= B_0^2 / \mu \rho_0(x)$ is the unperturbed Alfv\'en speed and $c_s^2(x) = \gamma p_0 / \rho_0 (x) $ is the unperturbed sound speed.  In equations (\ref{vxevolution} - \ref{vzevolution}) the nonlinear and nonideal terms can be found on the right-hand-side. Note that, equations (\ref{rho} - \ref{vzevolution}) reduce to the corresponding equations in Botha {{et al.}} (\cite{Botha2000}) in an ideal plasma.

Considering equation (\ref{vzevolution}) and neglecting the nonlinear and nonideal terms (right-hand-side), we see that ${{\rm{v}}_z}$ corresponds to the linear Alfv\'en wave. Thus, if ${\rm{v}}_A = {\rm{v}}_A(x)$, then both ${{\rm{v}}_z}$ and $B_z$ will depend on $x$. Hence, there will be coupling to both magnetoacoustic modes as follows: Firstly,  we can see that the first term in $N_2$ (i.e. $\sim{B_z{\partial B_z }/ {\partial x}}$) will generate a fast wave (i.e. drives ${{\rm{v}}_x}$ term) via phase mixing. Secondly, we see that the first term in $N_3$  (i.e. $\sim{B_z{\partial B_z }/ {\partial y}}$) corresponds to  the Alfv\'en-wave pressure gradient, or the {\emph{ponderomotive force}}, and that this (nonlinear) force generates a slow wave (i.e. drives ${{{\rm{v}}_y}}$ term) along the background magnetic field. Also note that these two terms we have described are quadratic in terms of $B_z$ and so will have an amplitude that is related to the square of the amplitude of the linear Alfv\'en wave.

%---------------------------------------------------------------

\subsection{Numerical Set-up}\label{section:2.3}

We solve MHD equations (\ref{MHDequations}) using the  {\emph{LARE2D}}  Lagrangian-remap code  (Arber {{et al.}} \cite{Arber}).

%\textit{This section is for non-dimensional equations and the non-dimensionalisation has not been mentioned. Is it better to keep everything dimensional? Then we could say that $A=0.01 {\rm{v}}_A(x=0)$ and $\omega = 0.25 {\rm{v}}_A(0)/L$ or whatever it is.}

We consider a numerical domain with $-100 \le x \le 100$, $0\le y \le 10,000$ with uniform-grid spacing in the $x-$direction and a stretched grid in the $y-$direction. The stretched grid places the majority of the $y-$direction gridpoints near low values of $y$. Results presented in this paper have a typical numerical resolution of $2,000 \times 20,000$ which means that $\delta x\approx \delta y\approx 0.1$ (due to the stretched grid). We chose such a large domain to ensure that no wave energy is reflected back into our system during our numerical simulations.

%\textit{Check this. With 20,000 points in y over a distance 10,000 suggests dy = 0.5}.

We drive our system with linearly-polarised, sinusoidal Alfv\'en waves. This harmonic wave train is driven at the $y=0$ boundary such that:
\begin{eqnarray}
{{\rm{v}}_z}(x,0,t)&=&A\sin{\left(\omega t\right)}\;, \;{{\rm{v}}_x}(x,0,t) = {{\rm{v}}_y}(x,0,t) = 0\;,\label{driven}\\
B_z(x,0,t)&=&-A\sqrt{\rho(x)} \sin{\left(\omega t\right)}\;,\;\left.{\frac{\partial B_x}{\partial y}}\right|_{y=0} = \left.{\frac{\partial B_y}{\partial y}}\right|_{y=0} = 0\;,\nonumber
\end{eqnarray}
where $A$ is the amplitude and $\omega$ is the frequency.  All other quantities have zero gradient boundary conditions at the $y-$boundaries, and we utilise periodic boundary conditions at the $x-$boundaries. In this paper, we set $A=0.01$ and $\omega=0.25$. At the start of the numerical run, the driven Alfv\'en wave is ramped up to $A$ over the first four wavelengths. {{It should be noted that boundary conditions (\ref{driven}) do not drive a pure Alfv\'en wave into the numerical domain; slow magnetoacoustic modes are nonlinearly excited as well with amplitude of order  $A^2$ (see $\S\ref{ideal1D}$).}}

{{Note that our choice of equilibrium (equation \ref{B_0}) is formally only an equilibrium in ideal MHD. However, numerical tests with no driving term show that this equilibrium is still valid over long timescales, i.e. the results that follow are due to the driven Alfv\'en waves, not due to our choice of equilibrium.}}

%----------------------------------------------------------------

\section{One-dimensional system}\label{section:1-D}

Let us start by considering the nonlinear and nonideal aspects of the Alfv\'en wave. In this section, we set $\rho_0 = \rm{constant}$, i.e. we remove the density inhomogeneity.  This effectively reduces equations (\ref{rho} - \ref{vzevolution}) to a one-dimensional (1D) system where variations in the  $x-$direction are ignorable (${{\partial} / {\partial x}} =0$). This 1D system will clearly illustrate the nonlinear driver in the equations as well as the effect of viscosity and resistivity. These effects must be identified before trying to interpret the results with a density inhomogeneity.

The driven, linear Alfv\'en wave has the form:
\begin{equation}
{{\rm{v}}_z} = {\mathcal{F}}(\omega [t - y/{\rm{v}}_A])\;, \quad B_z = -\sqrt{\mu \rho_0} {{\rm{v}}_z}\;,
\label{eq:linearAlfven}
\end{equation}
where the arbitrary function ${\mathcal{F}}$ can include a form for the ramp-up of the driver as well as the periodic driver. To illustrate the terms, we ignore the ramp-up and consider the sinusodial  driver; the solutions are:
\begin{eqnarray*}
{{\rm{v}}_z} &=& A \sin \left ( \omega [ t - y/{\rm{v}}_A]\right ) \quad {\rm{for}} \quad 0 < y < {\rm{v}}_A t\;,\\
B_z &=& - \sqrt{\mu \rho_0} A \sin \left ( \omega [t - y/{\rm{v}}_A]\right ) \quad {\rm{for}}\quad 0 < y < {\rm{v}}_A t\;,
\end{eqnarray*}
and
\begin{eqnarray*}
{{\rm{v}}_z} = B_z = 0 \quad {\rm{for}}\quad y > {\rm{v}}_A t\;.
\end{eqnarray*}

The nonlinear evolution of the Alfv\'en wave in an ideal, $\beta=0$ plasma can be found in Appendix \ref{appendix:beta=0}. However, this result can be easily derived from the $\beta \neq 0$ case (\S\ref{ideal1D}) and so Appendix \ref{appendix:beta=0} is only included for completeness.

%----------------------------------------------------------------

\begin{figure*}[t]
\begin{center}
\includegraphics[width=7.2in]{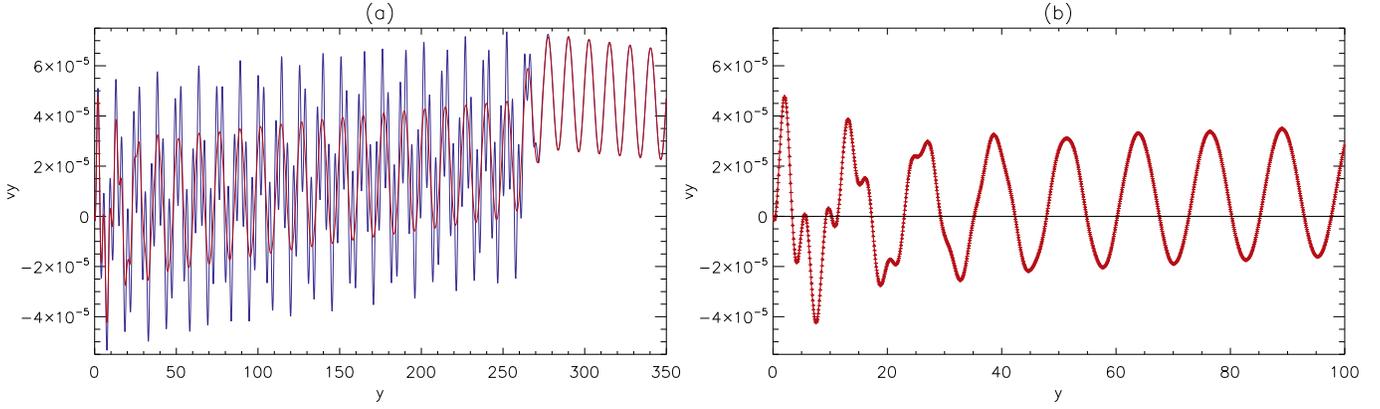}
\caption{$(a)$ Longitudinal perturbations ${{\rm{v}}_y}$ for $\eta=0.01$, $\beta=0.1$ plasma comparing $\eta=0.01$, $\nu=0$ (blue) and $\nu=0.01$, $\eta=0$ (red). Note only $0\le y\le 350$ is shown and that the agreement after $y\approx280$ is excellent (the two curves lie on each other). $(b)$ Blow-up of ${{\rm{v}}_y}$ over region $0 \le y \le 100$ for $\nu=0.01$, $\eta=0$ plasma only, where crosses indicate grid point resolution. Black line indicates ${\rm{v}}_y=0$. In both subfigures, $t=1000 \tau_A$.}
\label{figure4}
\end{center}
\end{figure*}

\subsection{Ideal $\beta > 0$ plasma}\label{ideal1D}

Consider a $\beta\neq0$ plasma, where we set $\beta=0.1$. The addition of finite-$\beta$ has no effect on the ${{\rm{v}}_z}$ component of the driven Alfv\'en wave (i.e. equation \ref{eq:linearAlfven} is still valid) but does influence the longitudinal component of velocity generated by the ponderomotive force. This can be seen in Figure \ref{figure2}a. Here, we see that there are two kinds of longitudinal motions in the system: the first can be seen between $y=c_s t \approx 280$ and $y= {\rm{v}}_A t=1000$. These longitudinal motions have a maximum amplitude of $A^2/4 \left( {\rm{v}}_A^2- c_s^2 \right)$ (this maximum is denoted by the horizontal blue line) and they are always positive. Thus, there is a net outflow.

The second type of longitudinal motion can be seen between  $y=0$ and $y=c_s t$. These longitudinal perturbations are acoustic waves (in 1D) with both positive and negative motions of the plasma along the field, and can be identified as a boundary-driven slow wave. Note that the longitudinal motions for $0 \le y \le c_s t$ are a combination of these two motions: the acoustic / slow wave (speed $c_s$) {\emph{and}} the ponderomotive perturbations (speed ${\rm{v}}_A$).

In Figure \ref{figure2}b, we can see a blow-up of the region $100\le x \le 200$ for ${{\rm{v}}_y}$ (black line). The red line represents an analytical solution describing both types of longitudinal motions (see equations \ref{alan solution_in_text} below). The agreement is excellent. Note that the slow wave has only propagated  a  distance of $y=c_st$ at $t=1000\tau_A$ and so, after  $y=c_st \approx 280$, only one wave is present.

The analytical solution for these nonlinearly-generated longitudinal motions is found by substituting equation (\ref{eq:linearAlfven}) into the weakly-nonlinear form of equation (\ref{vyevolution}), namely:
\begin{eqnarray*}
\left(\frac{\partial^2} {\partial t^2} - c_s^2\frac{\partial ^2}{\partial y^2} \right) {{\rm{v}}_y} = {{1} \over {\rho_0}}{{\partial N_3}\over {\partial t}} = -\frac{1}{\mu \rho_0}\frac{\partial^2}{\partial t\partial y}\left (\frac{B_z^2}{2}\right ).
\end{eqnarray*}
Thus, the analytical solution for ${{\rm{v}}_y}$ is:
\begin{eqnarray}
{{\rm{v}}_y} = \left \{\begin{array}{cc}
{\begin{array}{cc}\frac{A^2 {\rm{v}}_A}{4({\rm{v}}_A^2 - c_s^2)}\left (\cos [ 2 \omega (t - y/c_s)] \right.\\
\left.- \cos [2\omega (t - y/{\rm{v}}_A)] \right) \end{array}}   & 0 < y < c_s t\;,\\
\frac{A^2 {\rm{v}}_A}{4({\rm{v}}_A^2 - c_s^2)}\left ( 1 - \cos [2\omega (t - y/{\rm{v}}_A)] \right )& c_s t < y < {\rm{v}}_A t\;,\\
0 & y > {\rm{v}}_A t \;.
\end{array}\right.\label{alan solution_in_text}
\end{eqnarray}
which is valid only for early times (i.e. until nonlinear response becomes non-negligible). Note that these longitudinal motions were first reported in Botha et al. (\cite{Botha2000}).

%and ${{\rm{v}}_y} = 0$, for $y > {\rm{v}}_A t$.

Note that for $0 < y < c_s t$,  both types of wave in equation (\ref{alan solution_in_text}) {{must}} have the same frequency (i.e. twice the driving frequency) but will propagate at different speeds. Thus, they {\emph{must}} have different wavenumbers and hence different wavelengths. For the parameters chosen in this paper, these wavelengths are:
\begin{eqnarray*}
\lambda_{\rm{alfv{\acute{e}}n}} = 12.6\quad {\rm{and}}\quad \lambda_{\rm{slow}} = 3.63\;.
\end{eqnarray*}
Thus, for the typical numerical resolutions considered in this paper ($\delta y \sim 0.1$), both these perturbations are well resolved.

\begin{figure*}
\begin{center}
\includegraphics[width=7.2in]{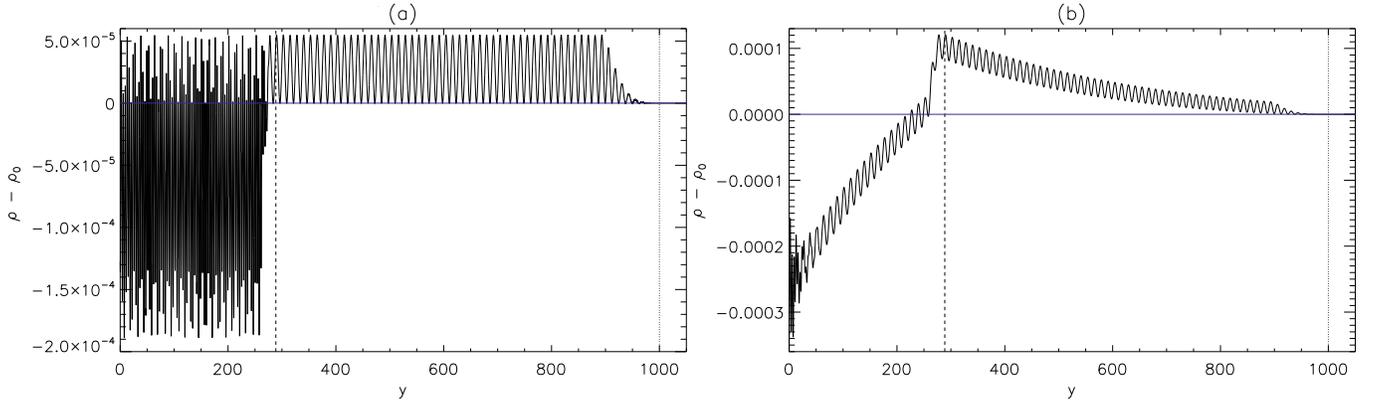}
\caption{$(a)$ Density evolution in ideal plasma. $(b)$  Density evolution in nonideal plasma ($\eta=\nu=0.01$). In both subfigures, $\beta=0.1$, $t=1000 \tau_A$, dashed (dotted) lines represent  $y=c_st$ ($y={\rm{v}}_At$) and the blue line denotes $\rho=\rho_0$.}
\label{figureDENSITY}
\end{center}
\end{figure*}

\subsection{Resistive plasma}\label{eta1D}

Let us now consider a resistive plasma, where we set $\eta = 0.01$ (and $\nu=0$, $\beta=0.1$). As before, an Alfv\'en wave is driven into the numerical domain  and the resultant ${{\rm{v}}_z}$ behaviour  can be seen in Figure \ref{figure3}a. It clear that the wave now experiences resistive damping. Such damping can be estimated from Fourier analysing the linear form of equation (\ref{vzevolution}) to give the dispersion relation:
\begin{eqnarray}
\omega^2 = \left( {\rm{v}}_A^2 + i\eta \omega \right) k^2\label{damping}
\end{eqnarray}
where $\omega=0.25$, ${\rm{v}}_A^2= {| {\bf{B}}_0 |}^2 /\rho_0$, $\eta=0.01$ and $k= k_R+ i k_I$. In Figure \ref{figure3}a, the dashed lines represent the envelope ${{\rm{v}}_z}= \pm A \exp{\left(k_I y\right)}$, where $k_I<0$. Assuming ${\rm{v}}_A^2 \gg \eta \omega$ (or equivalently that the magnetic Reynolds number is large) we can estimate that:
\begin{eqnarray}
k_I = - \frac{\eta \omega^2 }{2 {\rm{v}}_A^3} \label{ki}
\end{eqnarray}
to a high degree of accuracy.

Figure \ref{figure3}b shows the corresponding longitudinal motions (${{\rm{v}}_y}$) in our resistive system. As in Figure \ref{figure2}b, we see that the acoustic and ponderomotive wave components are again present, but that there is now a third phenomenon: a bulk flow in the positive $y-$direction with ${\rm{v}}_y>0$ and that has a maximum around $y=c_s t$.  Physically, this bulk flow (i.e. movement of material) is due to the ohmic heating in our system, and is a natural consequence of driving an Alfv\'en wave in a resistive plasma.

The longitudinal behaviour is governed by equation (\ref{vyevolution}), which in 1D (i.e. ${{\partial} / {\partial x}} =0$) and at early times (i.e. no nonlinear feedback) reduces to:
\begin{eqnarray}
&&\left( \frac{\partial^2  }{\partial t^2}  - c_s^2\frac{\partial^2  }{\partial y^2}\right) {{\rm{v}}_y}  =\nonumber \\
&&\qquad -{\frac{1}{ {\mu}{\rho_0}}}  \frac{\partial  }{\partial t} \left(  B_z {\frac{\partial B_z }{\partial y}} \right)  -{\frac{\left(\gamma-1\right) \eta }{ {\mu}{\rho_0}}}  \frac{\partial  }{\partial y} \left[ \left(  \frac{\partial B_z }{\partial y} \right) ^2 \right] \label{governingMAINTEXT}
\end{eqnarray}
The right-hand-side of this equation has two contributions: the first term depends upon the rate-of-change of the magnetic pressure gradient  ${\frac{\partial^2}{\partial t \partial y}}( {\frac{B_z^2}{2}}$) of the driven wave and the second depends upon pressure gradients created by the ohmic heating term (i.e. ${\frac{\partial}{\partial y}}  j_x^2  $) increasing the gas pressure. Note the second contribution vanishes under the ideal approximation.

Let us now evaluate these terms under the assumption that our linear Alfv\'en wave can be represented as:
\begin{eqnarray}
{{\rm{v}}_z} = A \sin { \left( \omega t -k_R y \right) } e^{ k_I y} \;.
\end{eqnarray}
Thus, from equations (\ref{v_z}) and (\ref{B_z}), $B_z$ has the form:
\begin{eqnarray}
B_z &=& -\frac{\omega B_0 A}{        {\rm{v}}_A^2  (k_R^2+k_I^2)} \left[ k_R \sin { \left( \omega t -k_R y \right) } \right.\nonumber\\
&&\qquad\left.-k_I \cos { \left( \omega t -k_R y \right) } \right]  e^{ k_I y} \;.\label{Bz_compare}
%B_z= -\frac{\omega B_0 A}{        {\rm{v}}_A^2  (k_R^2+k_I^2)} \left[ k_R \sin { \left( \omega t -k_R y \right) }-k_I \cos { \left( \omega t -k_R y \right) } \right]  e^{ k_I y} \;.\label{Bz_compare}
\end{eqnarray}
The derivation of equation (\ref{Bz_compare}) is given in Appendix \ref{AppendixB}.

Hence, the right-hand-side of equation (\ref{governingMAINTEXT}) can be simplified such that:
\begin{eqnarray*}
\left( \frac{\partial^2  }{\partial t^2}  - c_s^2\frac{\partial^2  }{\partial y^2}\right) {{\rm{v}}_y}  = \hat{P}+\hat{R}+\hat{BF}
\end{eqnarray*}
where the terms representing the Ponderomotive force, $\hat{P}$, the resistive heating force, $\hat{R}$, and the non-oscillatory resistive bulk flow force,  $\hat{BF}$, are given by:
\begin{eqnarray}
{\hat{P}} &=&   \frac{\omega^3 A^2}{ {\rm{v}}_A^2  (k_R^2+k_I^2)} \left[ {k_R  \cos { 2\left( \omega t -k_R y \right) } }\right.\nonumber\\
&&\qquad\left.{+k_I \sin {2 \left( \omega t -k_R y \right)} } \right]  e^{ 2 k_I y} \label{hatP}\\
{\hat{R}} &=& - \frac{\omega^2 A^2 \left( \gamma -1\right) \eta}{ {\rm{v}}_A^2} \left[ {k_R  \sin { 2\left( \omega t -k_R y \right) } }\right.\nonumber\\
&&\qquad\left.{ +k_I \cos {2 \left( \omega t -k_R y \right)} } \right]  e^{ 2 k_I y} \label{hatR}\\
{\hat{BF}} &=& -  \frac{\omega^2 A^2 \left( \gamma -1\right) \eta k_I}{ {\rm{v}}_A^2}  e^{ 2 k_I y}\label{hatBF}
\end{eqnarray}
which is valid for $c_st < y < {\rm{v}}_A t$. The derivation of these equations is given in Appendix \ref{AppendixB}.

Note that the amplitude of each  term is proportional to the amplitude squared of the driven Alfv\'en wave and that they decay with height since $k_I<0$. In addition, we note that $\hat{P}$ and $\hat{R}$ are trigonometric terms  with twice the driven frequency and twice the wavenumber. In the absence of dissipation ($\eta=0$ and so $k_I=0$) there is no decay and ${\hat{R}}$ and ${\hat{BF}}$ vanish.

We also note that ${\hat{BF}}$ is, critically,  non-oscillatory and it is this term that will create the bulk flow in the longitudinal direction.

% In Figure \ref{figure4}a, the dashed lines represent the envelope ${{\rm{v}}_z}= \pm A \exp{\left(k_I y\right)}$.

\begin{figure*}
\begin{center}
\includegraphics[width=7.2in]{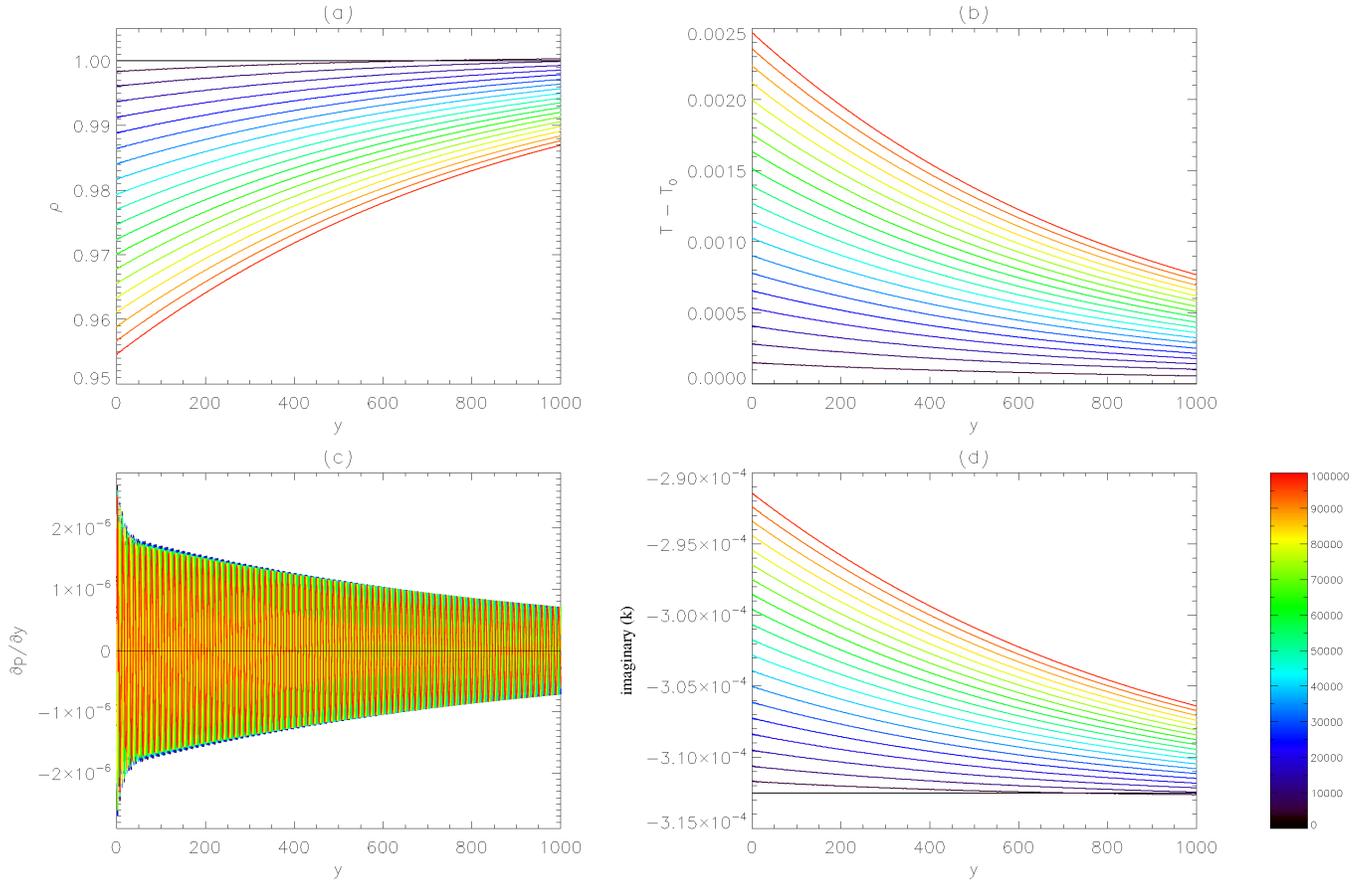}
\caption{Long-term evolution of $(a)$ density, $(b)$ temperature, $(c)$ pressure gradient (${\partial p}/{\partial y}$), and $(d)$ $k_I$, over  $0\le t\le 100,000 \tau_A$, where the different colours represent different times. The colour bar denotes the different times.}
\label{figure11}
\end{center}
\end{figure*}

\subsection{Viscous plasma}\label{nu1D}

Let us now consider a purely viscous plasma, where we set $\nu=0.01$, $\eta=0$ and $\beta=0.1$. Again, we drive an Alfv\'en wave into our numerical domain and the resultant ${{\rm{v}}_z}$ is identical to that seen Figure \ref{figure3}a. However, the damping mechanism is now due to  viscosity rather than resistivity, as in  $\S\ref{eta1D}$. Such damping can be estimated from Fourier analysing the linear form of equation (\ref{vzevolution}) to give the dispersion relation:
\begin{eqnarray}
\omega^2 = \left( {\rm{v}}_A^2 + i\nu \omega \right) k^2\label{visc_damping}\;\;.
\end{eqnarray}
Note the similarity to equation (\ref{damping}).

%$\nu=0.01$, $\eta=0$ (red)

Figure  \ref{figure4}a shows a comparison of ${{\rm{v}}_y}$ for the purely-resistive (blues, $\S\ref{eta1D}$, Figure \ref{figure3}b) and purely-viscous (red line) systems. Focusing on the viscous propagation, we see that three components are present: the ponderomotive wave  (between $280\le y \le 1000$), the bulk flow in the positive $y-$direction, and the acoustic perturbations (from $0\le y\le 280$). However, note that the acoustic perturbations now take a different form to those in the purely-resistive system (black line) and we note that in the viscous system the smaller wavelengths are rapidly damped out by viscosity. Above $y\approx 280$, the agreement is excellent (the two curves lie ontop of one {{another}}). Note that the oscillation is fully resolved, as can be seen in Figure \ref{figure4}b, which shows a blow-up of ${\rm{v}}_y$ in the region $0\le y\le 100$ in the viscous system.

In the viscous plasma, the bulk-flow phenomenon now comes from viscous heating, as opposed to ohmic heating as in $\S\ref{eta1D}$. Comparing the two systems (Figure  \ref{figure4}a) we see that the agreement for the ponderomotive component and bulk flow is excellent (unsurprising, since equations \ref{damping} and \ref{visc_damping} are identical for $\nu=\eta$). However, the viscous damping of the acoustic component is more pronounced (with an estimated damping length of $13.5$).

\section{Bulk-Flow Phenomenon}\label{section:bulk_flow}

Let us now investigate the effect that our longitudinal bulk flow  has on the equilibrium density profile, starting with the ideal case. Figure \ref{figureDENSITY}a  shows the evolution of the density perturbations at $t=1000\tau_A$ in our ideal plasma. We can clearly see that the contributions from the ponderomotive component and acoustic wave components. We also see that the density profile is perturbed about the equilibrium but that these perturbations do not grow in time.

Figure \ref{figureDENSITY}b shows the  density profile at $t=1000\tau_A$ in our visco-resistive ($\eta=\nu=0.01$) system. Here, in addition to the contributions from the ponderomotive component and acoustic wave, we see there is also a third component; i.e. the bulk flow has shifted the density profile in the positive $y-$direction, leading to a decrease in the density profile around $y=0$.

The physical interpretation of this bulk flow is as follows: visco-resistive heating increases the local temperature (with maximum around $y=0$). This increase in temperature increases the thermal pressure, and the resultant pressure gradient drives the bulk flow. Note that the bulk flow is not due to the  ponderomotive wave pressure force, otherwise it  would be apparent in our ideal plasma (i.e. in Figure \ref{figureDENSITY}a). Thus, the bulk flow is a natural consequence of driving an Alfv\'en wave in a nonideal plasma.

\begin{figure*}[t]
\begin{center}
\includegraphics[width=7.2in]{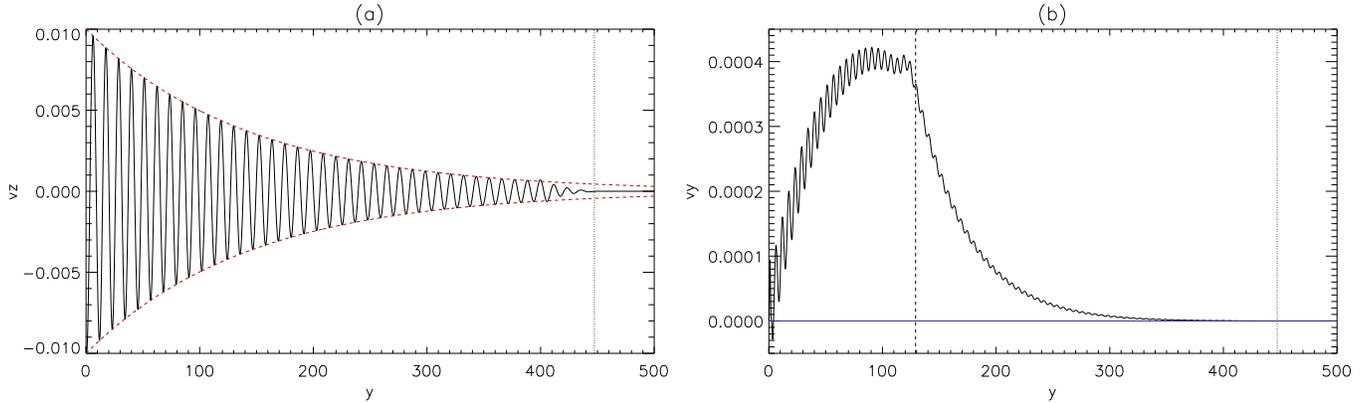}
\caption{$(a)$ Transverse perturbations ${{\rm{v}}_z}$ for $\eta=\nu=0.01$, $\beta=0.1$ plasma with $\rho_0=5$, where red dashed lines represent the envelope ${{\rm{v}}_z}= \pm A \exp{\left(k_I y\right)}$. $(b)$ Longitudinal perturbations ${{\rm{v}}_y}$ for same plasma, where black dashed line represents $y=c_s t$ and blue horizontal line denotes ${\rm{v}}_y=0$. In both subfigures, $t=1000\tau_A$ and  the vertical black dotted line represents $y={\rm{v}}_At$.}
\label{figure6}
\end{center}
\end{figure*}

%thus it is natural to ask whether the system will eventually reach a steady state.

Let us now look look at the long-term evolution of the density profile. Figure \ref{figure11}a shows the evolution of the density profile over $0\le t\le 100,000 \tau_A$, where the different colours represent different times (as denoted by the colour bar). It is clear that the density profile is modified over time, and at {\bf{$t=100,000 \tau_A$}} the density at $y=0$ is approximately $95.5\%$ of its equilibrium value. Figure \ref{figure11}b shows the evolution of the temperature profile, and we see that there is a steady increase in temperature over the whole evolution, with maximum increase at $y=0$. 

It is clear that the visco-resistive heating has substantially changed the background density profile, and that there is a substantial flow of plasma in the positive $y-$direction, causing a reduction in the density since the boundary conditions do not allow for this plasma to be replaced by an in-flow. As can be seen from Figure \ref{figure11}a, it appears that the rate-of-change (i.e. the decrease) in the density is decreasing (this is confirmed below in $\S\ref{section:4.1}$). There are two possible explanations for this decrease in the rate-of-change: either  the bulk flow (in the positive $y-$direction) has moved so much density that an opposing pressure gradient (in the negative  $y-$direction) has been set up, or it could be that at later times there is less visco-resistive heating in our system.

To examine the build-up of opposing pressure gradients, we can look at the evolution of ${\partial p}/{\partial y}$ in our system (where $p$ is total pressure), and this can be seen  in Figure \ref{figure11}c. We can see that there are both positive and negative perturbations, related to the oscillatory motions, but crucially  we find no evidence for the build-up of an opposing pressure gradient in the negative  $y-$direction.

%Thus, increases in temperature are associated with decreases in density, as we observe.

Let us now investigate the alternative hypothesis; at later times there is less  visco-resistive heating in our system. Recall that equation (\ref{hatBF})  highlighted the role of $k_I$ in our system; $k_I$ is relevant both to the damping of the driven Alfv\'en wave and to the enhanced bulk flow.  Figure \ref{figure11}d shows the evolution of $k_I$ in our system, and we see that the magnitude of $k_I$ is decreasing over time.  The behaviour of $k_I$ and the density profile are directly related by equation (\ref{ki}).

This provides the explanation for the decrease in the rate-of-change of the density profile: as the simulation proceeds, the visco-resistive damping heats the plasma, leading to a decrease in density. Thus (through equation \ref{ki}), the magnitude of $k_I$ decreases, thus there is less  damping in the system, and thus the rate of  visco-resistive  heating decreases. This in turn leads to a decrease in the rate-of-change of density. There is a strong feedback effect: a decrease in density leads to  a decrease in $|k_I|$ and thus a decrease in the nonideal heating, and thus a decrease in the bulk flow.

\begin{figure*}
\begin{center}
\includegraphics[width=7.2in]{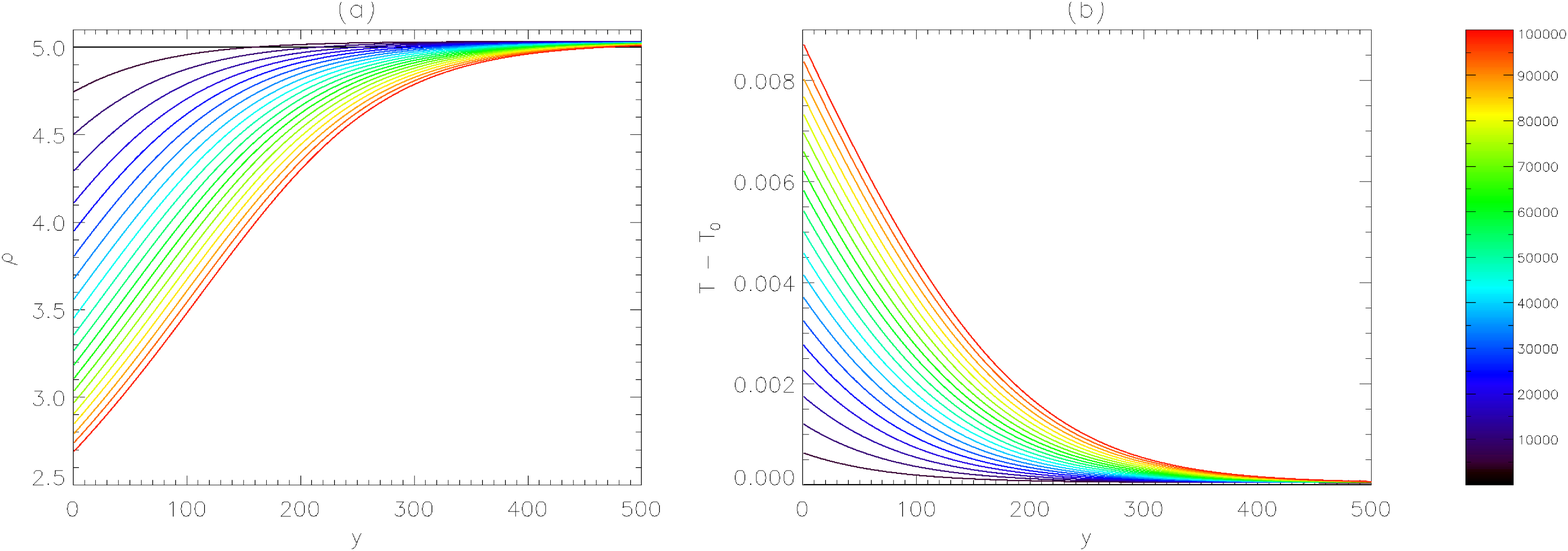}
\caption{Long-term evolution of$(a)$ density and $(b)$ temperature, over  $0\le t\le 100,000 \tau_A$, where the different colours represent different times. The colour bar denotes the different times.}
\label{figure14}
\end{center}
\end{figure*}

\begin{figure*}
\begin{center}
\includegraphics[width=7.2in]{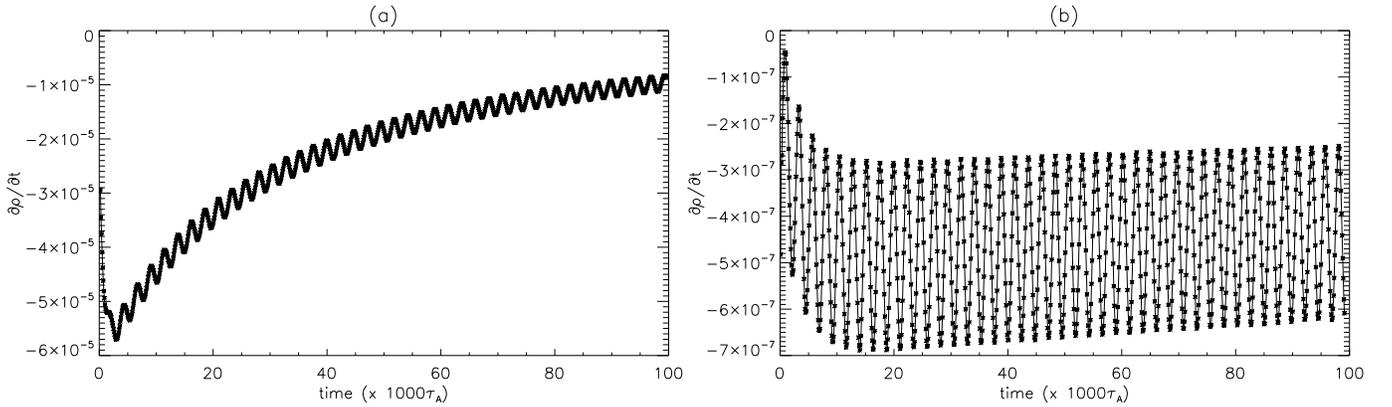}
\caption{Evolution of $\partial \rho / \partial t$ at $y=1$ for $(a)$ $\rho_0=5$ and $(b)$ $\rho_0=1$ plasmas.}
\label{figure12}
\end{center}
\end{figure*}

\begin{figure*}
\begin{center}
\includegraphics[width=7.2in]{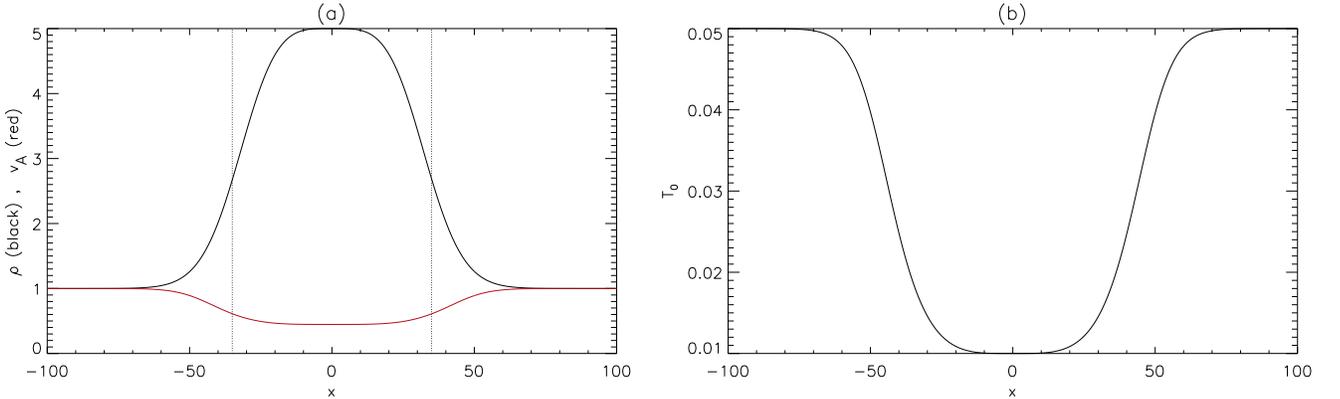}
\caption{$(a)$ Equilibrium density profile (black) and equilbrium Alfv\'en speed profile (red). $(b)$ Equilibrium temperature profile.}
\label{figure16}
\end{center}
\end{figure*}

\subsection{Dependence of equilibrium density profile}\label{section:4.1}

From equation (\ref{ki}) and the discussion in $\S\ref{section:bulk_flow}$, it is clear that  the larger the value of $|k_I|$ in our system, the more nonideal heating occurs at that time. Thus, there should be a strong dependence upon our choice of equilibrium density profile.

Let us consider a 1D system similar to that of $\S\ref{section:bulk_flow}$ (i.e. $\beta=0.1$, $\eta=\nu=0.01$, boundary conditions given by equation \ref{driven}) where we set $\rho_0=5$. Figure \ref{figure6}a shows the behaviour of ${{\rm{v}}_z}$ in this system at $t=1000\tau_A$. As in Figure \ref{figure3}a, we see that the driven Alfv\'en wave is damped. The magnitude of this resistive damping can be reproduced using dispersion relation (\ref{damping}), where we replace $\eta$ by $\eta+\nu$. Note that the wave is more rapidly damped than that in Figure \ref{figure3}a, since in the $\rho_0=5$ system there is a larger value of $|k_I|$ ($|k_I|=6.99\times 10^{-3}$ as opposed to $|k_I|=6.25\times 10^{-4}$ in the $\rho_0=1$ system). This can also be seen from equation (\ref{ki}).

Figure \ref{figure6}b shows the behaviour of ${{\rm{v}}_y}$ at $t=1000\tau_A$. As in Figure  \ref{figure4}a, we observe three components to the propagation: a ponderomotive component, an acoustic (boundary-driven) component and a monotonically-increasing quadratic component, i.e. the bulk flow in the positive $y-$direction, as described previously. Note that the bulk flow is significantly stronger than that examined in $\S\ref{section:bulk_flow}$. This is because in the $\rho_0=5$ system we have a larger value of  $|k_I|$ and thus more visco-resistive damping. Hence, there is more visco-resistive heating, which increases the local gas pressure by a greater degree, and it is this thermal-pressure gradient that drives the bulk flow.

Figures \ref{figure14}a and \ref{figure14}b show the long-term evolution of the density and temperature profile, respectively, in the $\rho_0=5$ system. As in the $\rho_0=1$ system, it is clear from Figure \ref{figure14}a that the density profile has decreased substantially during the simulation, and the density at $y=0$ is approximately $53.6\%$ its initial value after $t=100,000\tau_A$. Figure \ref{figure14}a also clearly shows the bulk flow of density in the positive $y-$direction. Figure \ref{figure14}b shows that there is a monotonic increase in temperature over the whole simulation with maximum increase at $y=0$.

%i.e. there is a propagation of a density pulse such than $\rho>\rho_0$.

Thus, it appears that a key variable in our system is the choice of equilibrium density profile: the greater the value of $\rho_0$, the greater the (initial) value of $|k_I|$ and hence the greater the magnitude of visco-resistive damping (which heats the plasma, which increases the thermal pressure, which drives the bulk flow due to pressure gradients). The bulk flow decreases the value of $\rho$, which decreases the value of  $|k_I|$, which reduces the amount of visco-resistive damping, and so on (strong feedback effect). Thus, we can understand the nature of the bulk-flow phenomenon over a range of density values.

Figures \ref{figure12}a and \ref{figure12}b show the rate-of-change of density (${\partial \rho} / {\partial t}$) at $y=1$ for the $\rho_0=5$ and $\rho_0=1$ systems, respectively. Let us first consider Figure \ref{figure12}a. it is clear that rate-of-change is decreasing as time evolves, i.e. $|{\partial \rho} / {\partial t}|$ is decreasing. Thus, as the simulation proceeds, density decreases at a slower rate, as expected from our explanation that as density decreases, there is less heating, leading to less thermal expansion. Figure \ref{figure12}b shows that, again, the rate-of-change of density is decreasing in the $\rho_0=1$ system but at a much slower rate than that seen in Figure \ref{figure12}a. Again, this is in agreement with our explanation.

%%%%%%%%%%%%%%%%%%%%%%%%%%%%%%%%%%%%%%%%%%%%%%%%%%%%%%%%%%%%%%%%%%%%%%%%%%%%%%%%%%%%%%%%%%

%----------------------------------------------------------------

\section{Two-dimensional inhomogeneous plasma}\label{section:2-D}

We now extend the model of $\S\ref{section:1-D}$ to include an inhomogeneous density profile. This will naturally introduce the phase mixing phenomenon into our system, in addition to the others  already discussed.

We consider the following equilibrium density profile:
\begin{eqnarray*}
\rho_0(x) = 1+4 {\rm{sech}}^2{\left[ \left( {\frac{x}{a}} \right)^2 \right]} \;,%\label{2D_density_profile}
\end{eqnarray*}
where, in this paper we set $a=35$. This equilibrium density profile  ($\:\rho_0$) can be seen in Figure \ref{figure16}a (black line); the red line represents the corresponding equilibrium Alfv\'en-speed profile. The equilibrium temperature profile ($T_0$) is shown in Figure \ref{figure16}b, which has been chosen such that the equilibrium gas pressure is constant.

%In order to keep the gas pressure constant, 

As detailed in $\S\ref{section:2.3}$, we solve MHD equations (\ref{MHDequations}) in a  numerical domain of $(x, y)$ $\in [-100, 100] \times [0,10^4]$ with a uniform / stretched grid  in the $x-$ / $y-$direction. We drive our system with linearly-polarised, sinusoidal Alfv\'en waves  at $y=0$ (boundary condition \ref{driven}). All other quantities have zero gradient boundary conditions at the $y-$boundaries and  we utilise periodic boundary conditions at the $x-$boundaries.

The numerical results can be considered over two times scales: short-term evolution, over which the classical phase mixing solution dominates, and the long-term evolution over which nonlinear effects can become important.

\begin{figure}
\begin{center}
\includegraphics[width=3.7in]{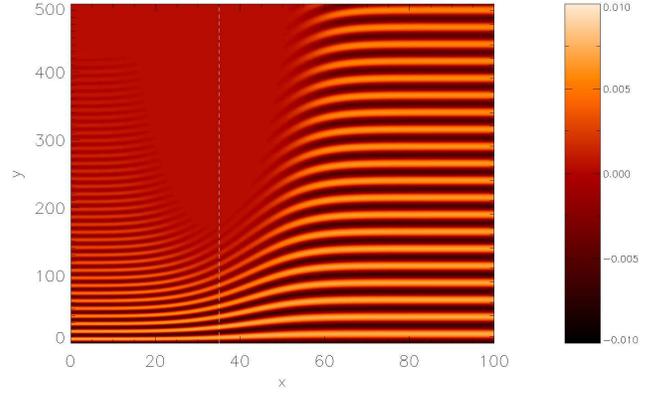}
\caption{Contours of ${{\rm{v}}_z}$ for $0\le x \le 100$, $0\le y \le 500$, at $t=1000\tau_A$. White dashed line denotes $x=35$. {{The temporal evolution is shown in the movie available in the on-line edition.}}}
\label{figure17}
\end{center}
\end{figure}

%Results is the we have drifting of heating layers - as well as bulk flow as described in section \ref{section:bulk_flow}. 2D works like this push, shear, heat, drift.

\subsection{Evolution of ${{\rm{v}}_z}$: Alfv\'en waves} %{Classic phase mixing solution}

Figure \ref{figure17} shows a contour plot of ${\rm{v}}_z$ for $0\le x \le 100$, $0\le y \le 500$ at time $t=1000\tau_A$. This contour plot shows the propagation of the Alfv\'en waves in our system, where the light and dark bands denote the peaks and troughs, respectively, of the wave motions in the $z-$direction. Three different behaviours are apparent: the oscillations along $x=0$ correspond to the propagation of the Alfv\'en wave in a $\rho_0=5$ plasma and a cut along $x=0$ reveals a profile identical to Figure \ref{figure6}a. Similarly, the oscillations along  $x=100$ correspond to the propagation of an  Alfv\'en wave in a $\rho_0=1$ plasma and a cut along $x=100$ is identical to that in Figure \ref{figure3}a. As expected, the propagation along $x=0$ (where $\rho_0[0,y]=5$) is damped much more rapidly than that along $x=100$ (where $\rho_0[100,y]=1$). Thus, the propagation along $x=0$ and $x=100$ and the (visco-resistive) damping mechanism  can be fully understood from the corresponding one-dimensional results.

\begin{figure}
\begin{center}
\includegraphics[width=3.6in]{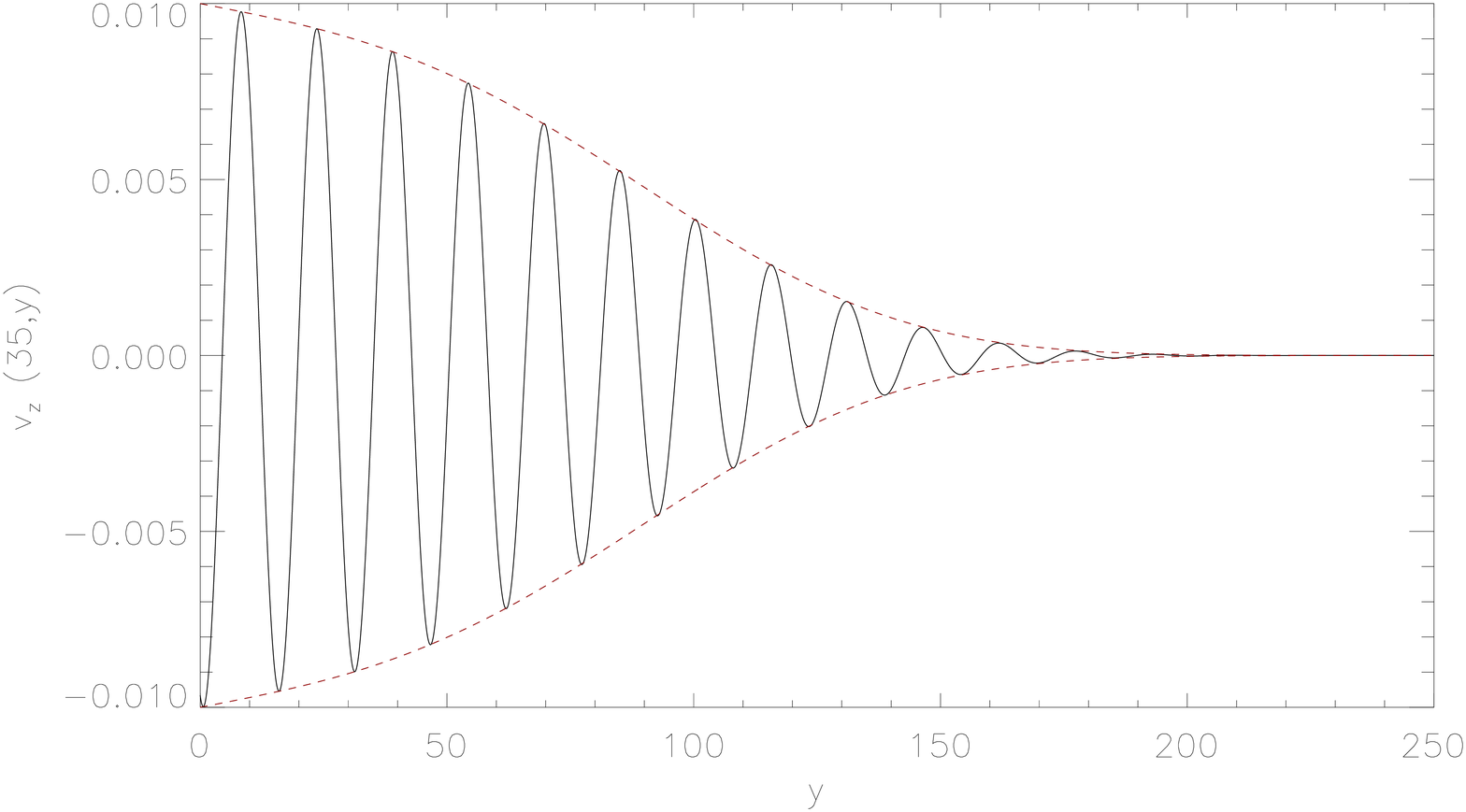}
\caption{Plot of ${{\rm{v}}_z}(35,y)$. Heyvaerts \& Priest solution overplotted}
\label{figure19}
\end{center}
\end{figure}

Let us now consider a cut along $x=35$ in Figure  \ref{figure17}, i.e. ${{\rm{v}}_z}(35,y)$; this can be seen in  Figure \ref{figure19}. Here, we see that the wave is rapidly damped (more rapidly than standard visco-resistive damping).

% and that the damping mechanism at work here is due to phase mixing.

The classical phase mixing solution of Heyvaerts \& Priest (\cite{Heyvaerts1983}) is given by:
\begin{eqnarray}
{{\rm{v}}_z} = \pm A \exp{\left(k_I y\right)} \cdot \exp{ \left[ -\frac{1}{6}\left( \frac{k_\parallel y}{R^{1/3}}\right)^3\right]} \;\label{HP1883_formula}\;,
\end{eqnarray}
where $k_\parallel={\omega}/{{\rm{v}}_A(x)}$ and $R={\omega}/{ (\eta+\nu) {\left[ {\frac{d}{dx} \left( \ln{k_\parallel}  \right) }  \right]}^2 }$. This solution is overplotted in Figure \ref{figure19} (red dashed line) and the agreement is excellent. Thus, at early times, the damping of ${{\rm{v}}_z}$ is well understood as that corresponding to the  classic phase mixing mechanism.

A movie of the evolution of ${{\rm{v}}_z}$ over the whole simulation (up to  $t=100,000\tau_A$) is associated with Figure \ref{figure17}. It is clear that the behaviour of ${{\rm{v}}_z}$ changes very little over the whole simulation, and that Alfv\'en waves are continuously driven into the numerical domain; there is no evidence of the system choking itself off. Note that the small changes in ${{\rm{v}}_z}$ are dictated by the changes in the density profile (see $\S\ref{section:density2D}$ below).

\begin{figure}
%\begin{center}
\includegraphics[width=3.6in]{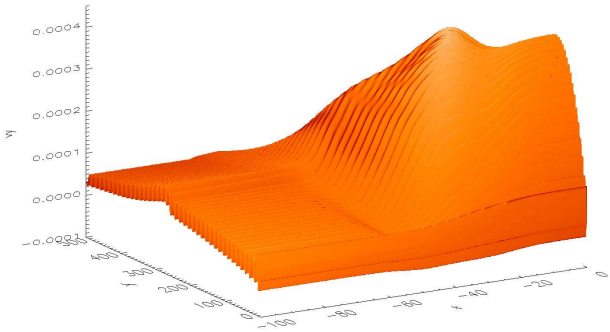}
\caption{Shaded surface of ${{\rm{v}}_y}(x,y)$ at $t=1000\tau_A$. {{The temporal evolution is shown in the movie available in the on-line edition.}} }
\label{figureVY}
%\end{center}
\end{figure}

\vspace{5cm}

\subsection{Evolution of ${{\rm{v}}_y}$: Slow magnetoacoustic waves}

Figure \ref{figureVY} shows a shaded-surface of ${{\rm{v}}_y}(x,y)$ at $t=1000\tau_A$. As in the analysis of ${{\rm{v}}_z}$, a cut along $x=0$ reveals behaviour identical to that of  Figure \ref{figure6}b, and a cut along $x=100$ reveals behaviour identical to that of ${{\rm{v}}_y}$ in a one-dimensional plasma with $\beta=0.1$, $\eta=\nu=0.01$ at $t=1000\tau_A$, i.e. similar to Figure \ref{figure4}a. Thus, as in our one-dimensional analyses, we see that our simulation contains three types of longitudinal motions: slow magnetoacoustic waves (previously called acoustic waves in $\S\ref{section:1-D}$), ponderomotive wave components (propagating at the Alfv\'en speed) and a bulk flow in the positive $y-$direction. As expected, all amplitudes are of the order $A^2$.

A movie of the evolution of ${{\rm{v}}_y}$ over the whole simulation (up to  $t=100,000\tau_A$) is associated with   Figure \ref{figureVY}. It is clear that the growth of ${{\rm{v}}_y}$ saturates, and that the behaviour of ${{\rm{v}}_y}$ is dictated by that of ${{\rm{v}}_z}$ and the heating profile. Thus, the complete evolution of ${{\rm{v}}_y}$ is of secondary importance in understanding the evolution of the density and temperature profiles.

%VY behaviour $\Rightarrow$ SLOW waves, as expected from 1D. SATURATES (so not critically important).

\begin{figure*}
\begin{center}
\includegraphics[width=7.2in]{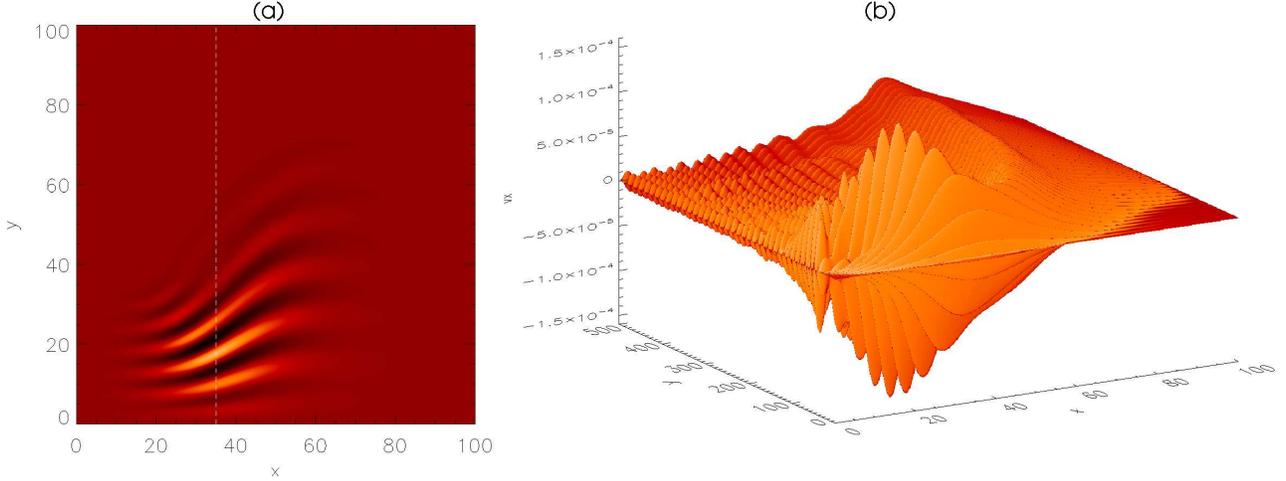}
\caption{$(a)$ Contour of ${{\rm{v}}_x}(x,y)$ at $t=100\tau_A$. $(b)$ Shaded surface of ${{\rm{v}}_x}(x,y)$ at $t=1000\tau_A$.   {{The temporal evolution of the right panel is shown in the movie available in the on-line edition.}} } 
\label{figureVX}
\end{center}
\end{figure*}

%%%%%%%%%%%%%%%%%%%%%%%%%%%%%%%%%%%%%%%%%%%%%%%%%%%%%%%%%%%%%%%%%%%%%%%%%%%%%%%%%%

\subsection{Evolution of ${{\rm{v}}_x}$: Fast magnetoacoustic waves}

%Here, we can see that oscillations in ${{\rm{v}}_x}$ have been generated,

In addition to Alfv\'en waves and slow magnetoacoustic waves, there is now a third type of MHD wave present in our system: fast magnetoacoustic waves, which can be seen in Figure \ref{figureVX}. These fast waves are nonlinearly generated by transverse gradients in the Alfv\'en-speed profile, in agreement with the analytical work of Nakariakov {{et al.}} (\cite{Nakariakov1997}). The fast waves are permanently generated by phase mixing, and are refracted towards regions of lower Alfv\'en speed. This can be clearly seen in  Figure \ref{figureVX}a, which shows a contour plot of ${\rm{v}}_x$  behaviour at  $t=100\tau_A$. These oscillations have a maximum around $x=35$ and are refracted in towards $x=0$ (i.e. towards regions of lower Alfv\'en speed, also see Figure \ref{figure16}a). These wave motions are generated with a frequency twice that of the driving frequency, which is in agreement with previous analytical predictions (Nakariakov {{et al.}} \cite{Nakariakov1997};  \cite{Nakariakov1998}).

A shaded-surface of the behaviour of ${{\rm{v}}_x}$ at  $t=1000\tau_A$ can be seen in Figure \ref{figureVX}b. Here, we can see that the generated fast waves have propagated across the magnetic fieldlines and have set-up an interference pattern. The interference pattern also results from waves generated along $x=35$ overlapping with those generated along $x=-35$, and there is also overlap due to the periodic boundary conditions. However, as first noted by Botha {{et al.}} (\cite{Botha2000}), these fast waves grow initially and then saturate. Botha {{et al.}} explain this saturation in terms of simple wave interference: physically, the saturation of fast waves is due to destructive interference from incoherent sources.

A movie of the evolution of ${{\rm{v}}_x}$ over the whole simulation (up to  $t=100,000\tau_A$) is associated with   Figure \ref{figureVX}. As with the evolution of ${{\rm{v}}_y}$, it is clear that ${{\rm{v}}_x}$ quickly saturates at amplitudes of order $A^2$ and, hence, it is only of secondary importance in understanding the evolution of the density and temperature profiles.

\begin{figure*}
\begin{center}
\includegraphics[width=7.25in]{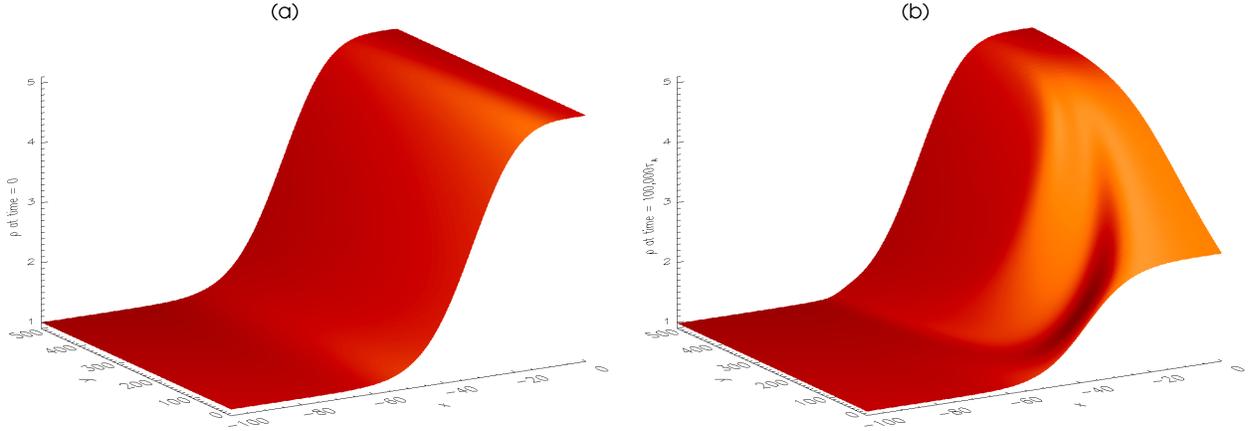}
\caption{$(a)$ Shaded surface of $\rho(x,y)$ at $t=0$ (i.e. equilibrium density profile). $(b)$ Shaded surface of $\rho(x,y)$ at $t=100,000\tau_A$.} 
\label{figure_final_density_profile_SHADESURF}
\end{center}
\end{figure*}

\begin{figure}
\begin{center}
\includegraphics[width=3.6in]{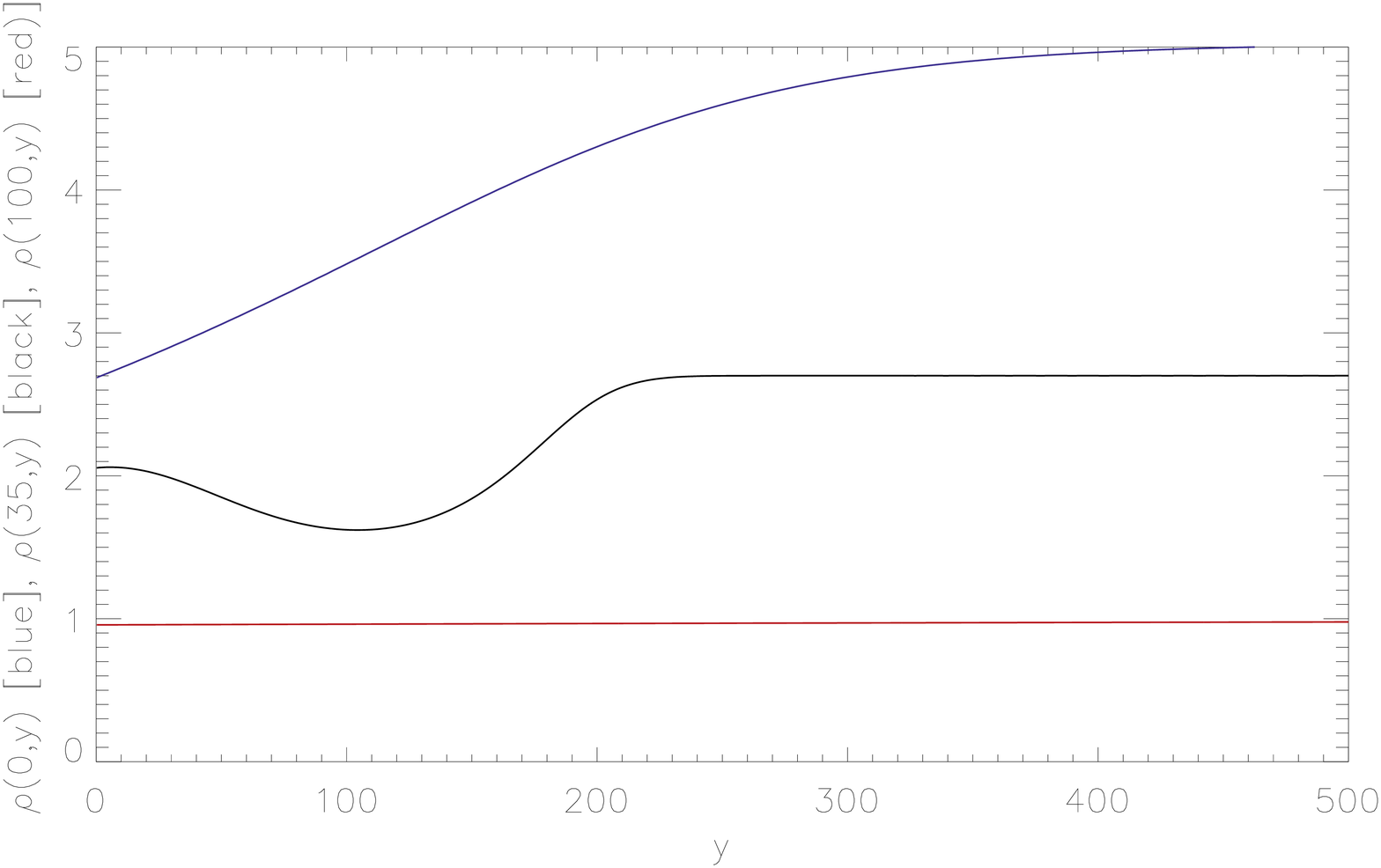}
\caption{Density profiles of $\rho(x,y)$ along the lines $x=0$ (blue), $x=35$ (black) and $x=100$ (red), at $t=100,000\tau_A$.} 
\label{figure_final_density_profile_cuts}
\end{center}
\end{figure}

\subsection{Evolution of density profile}\label{section:density2D}

Let us now look at the evolution of the density profile in our system. Figure \ref{figure_final_density_profile_SHADESURF} compares shaded-surfaces of the initial density profile (Figure \ref{figure_final_density_profile_SHADESURF}a) at $t=0$ and the final density profile (Figure \ref{figure_final_density_profile_SHADESURF}b) at  $t=100,000\tau_A$ at the end of our simulation. Note that we have presented $-100 \le x \le 0$ here, for clarity, and recall that the simulation is symmetric about $x=0$.

It is clear that the density profile has changed substantially during the simulation.  Figure \ref{figure_final_density_profile_cuts} shows cuts along $x=0$, $35$ and $100$ at the end of the simulation. The cuts along $x=0$ and $x=100$ are in excellent agreement with those predicted by the one-dimensional model (i.e. compare to Figures \ref{figure11}a and  \ref{figure14}a at $t=100,000\tau_A$). Thus, the nature of the density change at these locations can be understood in terms of the bulk-flow phenomenon described in $\S\ref{section:bulk_flow}$.

The nature of the density change along $x=35$ is however completely different. Here, the bulk flow has been suppressed (very little density change can be seen after $y\approx 225$) and instead the localised decrease in density is due to the strong, localised heating from phase mixing (see below).

%Also, this localised heating is free to move, meaning that a cut along $x=35 is not the best representation (moves in and out)$ 

\subsection{Evolution of temperature profile}

%Figure \ref{figure_final_density_profile_SHADESURF} compares the initial and final density profiles.
%Note that we have presented $-100 \le x \le 0$ here, for clarity (recall that the simulation is symmetric about $x=0$).

\begin{figure*}
\begin{center}
\includegraphics[width=7.2in]{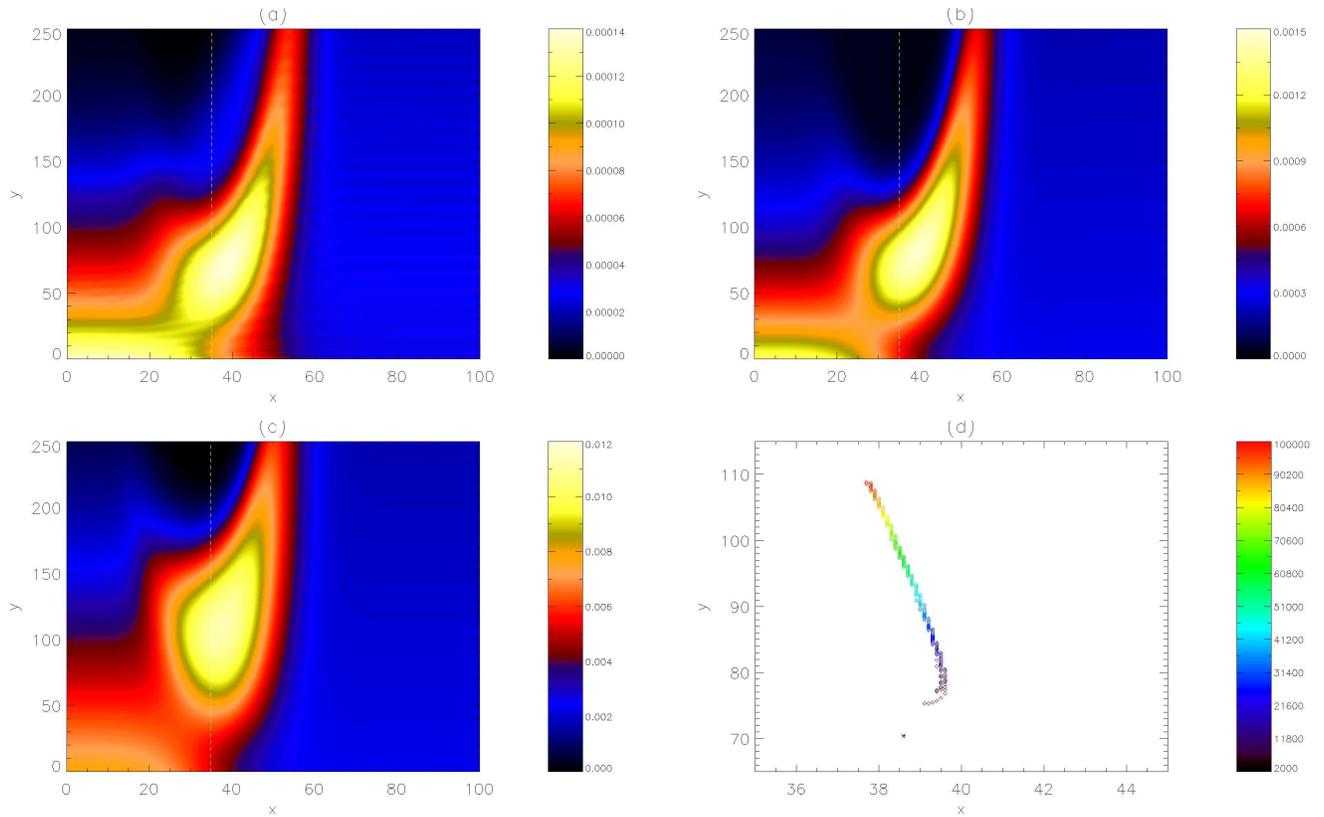}
\caption{Contour of $T - T_0$ at $(a)$ $t=1000\tau_A$, $(b)$ $t=10,000\tau_A$ and $(c)$ $t=100,000\tau_A$. Dashed white line denotes $x=35$. $(d)$ Plots (diamonds) the location of the maximum of $T - T_0$ over $2000\tau_A \le t \le 100,000\tau_A$. Also plotted (single star) is the location at $t=1000\tau_A$. In subfigures $(a) - (c)$, each colour bar represents magnitude, whereas in $(d)$ the corresponding colour bar donotes time.}
\label{figure_final_temperature_profile_contour}
\end{center}
\end{figure*}

Figure \ref{figure_final_temperature_profile_contour} shows contour plots of $T - T_0$ at times  $(a)$ $t=1000\tau_A$, $(b)$ $t=10,000\tau_A$ and $(c)$ $t=100,000\tau_A$. In Figure \ref{figure_final_temperature_profile_contour}a, we see that there is strong, localised heating (maximum occurs at $x=38.6$, $y=70.4$) which results  from the enhanced visco-resistive dissipation associated with phase mixing, producing a distinctive teardrop shape. Independently, the heating in the bottom left-corner is the visco-resistive heating associated Alfv\'en-wave damping  along $\rho_0=5$.

At a later time (Figure \ref{figure_final_temperature_profile_contour}b, $t=10,000\tau_A$) the overall temperature profile appears qualitatively similar to that in  Figure \ref{figure_final_temperature_profile_contour}a, i.e. phase mixing is still the dominant heating mechanism. However, the maximum of $T-T_0$ (located at $x=39.6$, $y=80.5$)  is now approximately ten times greater (note that the colour bar has changed).

Figure \ref{figure_final_temperature_profile_contour}c shows the temperature profile at the end of our simulation ($t=100,000\tau_A$). Again, qualitatively the contours appears similar to those in $(a)$ and $(b)$. However, the magnitude has again increased substantially (the colour bar has again changed) and the maximum of $T-T_0$ is now located at $x=37.7$, $y=108.8$. Again, phase mixing is still the dominant heating mechanism.

% and there is evidence that the location of maximum heating has moved.
%The colour bar has again changed and the maximum of $T-T_0$ is now located at $x=37.7$, $y=108.8$. Thus, phase mixing is still the dominant heating mechanism and there is evidence that the location of maximum heating has moved.

Figure \ref{figure_final_temperature_profile_contour}d shows the location of the maximum of $T-T_0$ for $2000\tau_A \le t \le 100,000\tau_A$ (note the different length scales on the axes). It is clear that the location of the maximum of  $T-T_0$ changes during our simulation, and that this drifting is in the direction of decreasing Alfv\'en-speed profile. This result addresses one of the fundamental question asked in $\S\ref{section1}$, i.e. we find that drifting of the heating layer in nonideal, nonlinear simulations of phase mixing {\it{is}} possible. However, we have also shown  that such drifting is very small and takes a very long time, at least for the parameters considered in this paper.

Finally, note that before $t=2000\tau$, the maximum of  $T-T_0$ sometimes occurs due to the bulk-flow heating, rather than due to phase mixing (phase mixing takes a finite amount of time to become the dominant heating mechanism) and thus, for clarity, we have not included these points in Figure \ref{figure_final_temperature_profile_contour}d. However, for comparison we have included the location of the maximum of  $T-T_0$ at $t=1000\tau$ (i.e.  $x=38.6$, $y=70.4$) and this is denoted by a star.

%Opposing pressure gradients act to restore the pressure balance, which results in a decrease in the local density.

%----------------------------------------------------------------

\section{Conclusions}\label{section:conclusions}

We have investigated the nonlinear, nonideal behaviour of Alfv\'en wave propagation and phase mixing within an inhomogeneous environment, over long timescales. The governing MHD equations have been solved in 1D and 2D environments using both analytical techniques and numerical simulations. 

In an ideal, one-dimensional study (no density inhomogeneity, $\partial/\partial x=0$) we find that by driving a linear Alfv\'en wave (in ${\rm{v}}_z$) into our numerical domain, we also generate two types of longitudinal wave: boundary-driven acoustic waves (propagating at speed $c_s$) and a nonlinear perturbation (propagating at ${\rm{v}}_A$) which is driven by the ponderomotive force. Both these motions have a frequency twice that of the driven Alfv\'en wave, and an exact mathematical solution for both wave types was derived. The acoustic wave is degenerate under the $\beta=0$ approximation (see Appendix $\ref{appendix:beta=0}$) but the ponderomotive wave is always present in a nonlinear system.

We find that the addition of resistive and viscous effects naturally leads to visco-resistive damping (through the dispersion relation) and thus to visco-resistive heating, which in turn leads to the introduction of a new phenomenon: a bulk flow in the positive $y-$direction. Physically, this bulk flow is a direct response to the  visco-resistive heating in the system: the heating increases the local temperature which increases the local thermal pressure. The resulting presure gradient drives the bulk flow. The bulk flow is purely in the positive $y-$direction because the boundary conditions are held fixed (${{\rm{v}}_y}=0$). As a result of this out-flow, a  significant reduction in density occurs since the boundary conditions do not allow for the plasma to be replaced by an in-flow.  Note that the bulk flow is not due to the ponderomotive force, otherwise such a flow would be apparent in the ideal, nonlinear system. 

We find that the magnitude of the visco-resistive heating is strongly dependent upon $k_I$ and thus on our choice of $\rho_0$; the greater the value of $\rho_0$,  the more visco-resistive heating that occurs. More visco-resistive heating leads to a  stronger bulk flow in the longitudinal direction.

We also investigated the effect of driving an Alfv\'en wave in a nonideal plasma over very long timescales. We find that the equilibrium density profile changes substantially over the duration of our simulation (end of the simulation at $t=100,000\tau_A$) specifically due to this bulk-flow phenomenon (which is itself directly due to the visco-resistive heating). For a $\rho_0=1$ plasma, we find a decrease in density of about $4.6\%$ at $y=0$ but for a $\rho_0=5$ plasma we find a decrease of  about $46.4\%$. We also find that the rate-of-change of density at $y=0$ is decreasing as the simulation proceeds. This is because as the density decreases, the value of $|k_I|$ also decreases. This smaller value of $|k_I|$  reduces the amount of visco-resistive damping, which slows the rate-of-change of density, and so on. Thus, there is a strong feedback effect in our system, and we can explain the nature of the bulk flow and change of density profile over a range of density values. We conclude that the bulk-flow phenomenon is {\emph{a natural consequence of driving an Alfv\'en wave in a nonideal plasma}}.

%%%%%%%%%%%%%%%%%%%%%%%%%%%%%%%%%%%%%%%%%%%%%%

We then extended our analysis to  include an inhomogeneous density profile in a two-dimensional system and, as before, we drive our system with sinusodial, linear Alfv\'en waves. As expected from Heyvaerts \& Priest (\cite{Heyvaerts1983}), our simulation displays classic phase mixing. A cut showing ${{\rm{v}}_z}$ as a function of height along $x=35$ provided an excellent match with the analytical predictions of Heyvaerts \& Priest (\cite{Heyvaerts1983}). Cuts along $x=0$ ($\:\rho_0=5$) and $x=100$  ($\:\rho_0=1$) were identical to those of the corresponding visco-resistive damped Alfv\'en waves  seen in our one-dimensional analysis.

Our one-dimensional analysis also explained the longitudinal motions  (${{\rm{v}}_y}$) in our system and we again observed the boundary-driven acoustic waves (interpreted as slow magnetoacoustic waves in 2D, with speed $c_s$), our ponderomotive waves (propagating at ${{\rm{v}}_A}$)  and the bulk-flow phenomenon.

In addition to Alfv\'en waves and slow magnetoacoustic waves, our 2D system also contains fast magnetoacoustic waves (seen primarily in ${{\rm{v}}_x}$). These fast waves are  continuously generated by Alfv\'en-wave phase mixing at a frequency twice that of the driven Alfv\'en wave, and propagate across magnetic fieldlines and away from the phase mixing layer (as predicted by Nakariakov {{et al.}} \cite{Nakariakov1997}; \cite{Nakariakov1998}). These waves are also refracted towards regions of lower Alfv\'en speed.

Since there is a permanent leakage of energy away from the phase mixing layer, it is possible that these fast waves can cause indirect heating of the plasma as they propagate away and dissipate far from the phase mixing layer itself, thus spreading the heating across the domain (Nakariakov {{et al.}} \cite{Nakariakov1997}). However, due to our choice of amplitude ($A=0.01$) we find that only a small fraction of the Alfv\'en-wave energy is converted into fast waves and, thus, only a small amount of indirect heating occurs.

Over long timescales, we find that both the slow and fast waves saturate, and always remain at amplitudes of the order $A^2$. Hence, these waves only play a secondary role in the heating of the plasma.

We find that at the end of our simulation, the density profile has changed substantially. This is due to two effects: firstly, the bulk-flow phenomenon is naturally present in our system, with the density changes as predicted by our one-dimensional analysis. Secondly, there is a substantial decrease in density at the locations of phase mixing heating.

The change in the background density profile does not affect the propagation of the Alfv\'en wave (amplitude of the order $A$) to a great degree, and the behaviour of ${{\rm{v}}_x}$ (fast waves) and ${{\rm{v}}_y}$ (slow waves) do not have a strong feedback effect on the system (both amplitudes of the order $A^2$) and only play secondary roles in the system (i.e. $A=0.01$, so feedback effect is only of the order of $1\%$). However, these conclusions may be different for Alfv\'en waves driven with a much larger amplitudes.

%%%%%%%%%%%%%%%%%%%%%%%%%%%%%%%%%%%%%%%%%%%%%%

We have also investigated how the temperature profile changes during our simulation. We find a substantial increase in localised temperature, generated by the phase mixing mechanism, over the duration of our simulation, and find that the maximum temperature increases by a factor of about a hundred during the simulation. We also find that the location of the maximum temperature drifts during our simulation, thus providing an answer to one of the key questions asked in  $\S\ref{section1}$, i.e. we find that drifting of the heating layer {\it{is}} possible in a nonlinear phase mixing scenario.

However, we also find that this drifting is very small and occurs over a very long timescale (our simulation runs over $100,000\tau_A$),  at least for the parameters we have considered in this paper. For typical coronal parameters, $\tau_A=60$ seconds and so we are considering timescales of approximately $69.4$ days, which is too long to be physically important in the corona (i.e. other coronal processes act over much shorter timescales). However, it may be possible to increase the speed and magnitude of this drifting by significantly increasing the magnitude of heating in our system, i.e. attempt to increase the amount of enhanced dissipation from phase mixing. Thus, future studies could consider increasing the amplitude and frequency of the driven Alfv\'en wave, or increasing the steepness of the gradient in our background density inhomogeneity.

%This work has provided evidence that the drifting of the heating layer is possible in a phase mixing scenario. However, we also find that this drifting is very small and occurs over a very long timescales (our simulation runs over $100,000\tau_A$). 

%----------------------------------------------------------------

\begin{acknowledgements}
JAM and IDM acknowledge financial assistance from the Leverhulme Trust and Royal Society, respectively. JAM acknowledges IDL support provided by STFC. JAM also wishes to thank Valery Nakariakov, Tony Arber and Hugh Hudson for helpful and insightful discussions and suggestions. The computational work for this paper was carried out on the joint STFC and SFC (SRIF) funded cluster at the University of St Andrews (Scotland, UK).
\end{acknowledgements}

%----------------------------------------------------------------

%----------------------------------------------------------------

\begin{appendix}

%----------------------------------------------------------------

\section{Alfv\'en wave propagation in a one-dimensional,  ideal, $\beta=0$ plasma}\label{appendix:beta=0}

\begin{figure*}
\begin{center}
\includegraphics[width=7.0in]{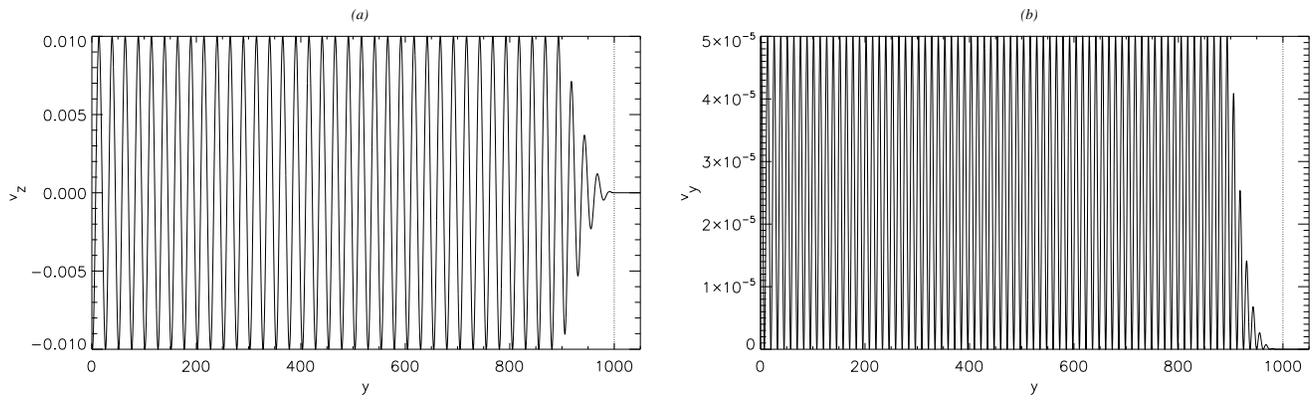}
\caption{$(a)$ Transverse perturbations ${{\rm{v}}_z}$ for ideal ($\eta=\nu=0$) $\beta=0$ plasma. $(b)$ Longitudinal perturbations ${{\rm{v}}_y}$ for ideal  $\beta=0$ plasma. In both subfigures, $t=1000\tau_A$ and  dotted line represents $y={\rm{v}}_At$.}
\label{figure_appendix}
\end{center}
\end{figure*}

Consider a cold $(\beta=0$), ideal $(\eta = \nu =0)$ plasma in our one-dimensional system ($\S\ref{section:1-D}$). Driving this system with boundary condition (\ref{driven}) generates an Alfv\'en wave at $y=0$; this can be seen in Figure \ref{figure_appendix}a. The Alfv\'en wave (${{\rm{v}}_z}$) propagates in the direction of increasing/positive $y$ with amplitude $A=0.01$ and there is no damping. The time of the snapshot can be read directly from the $y-$axis using the relation $t=y/{\rm{v}}_A$. The ramp-up over the first four wavelengths is clearly visible.

Figure \ref{figure_appendix}b shows the longitudinal motions (${{\rm{v}}_y}$) in the system. Here, ${{\rm{v}}_y}$ is driven by  the nonlinear terms of equation (\ref{vyevolution}) ($\sim  B_z {{\partial B_z }/ {\partial y}}$) and again there is no damping. In this paper, we call this second-order nonlinear effect the ponderomotive effect, and thus these longitudinal motions, which propagate at the Alfv\'en speed, are driven by the ponderomotive force (Alfv\'en wave-pressure gradients). At early times, these longitudinal motions are governed by a simplified version of equation (\ref{vyevolution}):
\begin{eqnarray}
{{\partial^2  {{\rm{v}}_y}}\over {\partial t^2}}  = -{{1}\over{ {\mu}{\rho_0}}} {{\partial }\over {\partial t}} \left( B_z {{\partial B_z }\over {\partial y}} \right) \;.\label{SLOW_early_cold_ideal}
\end{eqnarray}
Assuming  boundary conditions  (\ref{driven}) in our ideal, cold plasma and that  ${{\rm{v}}_x}$, ${{\rm{v}}_y}$ are initially zero,  equation (\ref{SLOW_early_cold_ideal}) has an analytical solution of the form:
\begin{eqnarray}
{{\rm{v}}_y} = \frac{A^2}{4{\rm{v}}_A}\left\{ 1 -   \cos {\left(2\omega [t - y/{\rm{v}}_A] \right)} \right\} \;,\label{agreement}
\end{eqnarray}
which is valid for $0<y<{\rm{v}}_A t$. Note that these perturbations  have an amplitude of $\mathcal{O}(A^2)$ and are always positive for an Alfv\'en wave propagating in the positive $y-$direction (due to the constant of integration, and thus due to the boundary conditions). Finally, note that the frequency of these longitudinal motions is twice that of the driving frequency, and that the motions do not grow with time. Interestingly, the ramp-up to maximum amplitude now occurs over eight wavelengths, i.e. twice that of the driven wave.

%----------------------------------------------------------------

\section{Derivation of equations \ref{Bz_compare} - \ref{hatBF}}\label{AppendixB}

Equation (\ref{vyevolution}) governs the driven, longitudinal motions in the system, and in 1D (${\partial}/{\partial x}=0$) has the form:
\begin{eqnarray}
&&\left( \frac{\partial^2  }{\partial t^2}  - c_s^2\frac{\partial^2  }{\partial y^2}\right) {{\rm{v}}_y}  =\nonumber \\
&&\qquad -{\frac{1}{ {\mu}{\rho_0}}}  \frac{\partial  }{\partial t} \left(  B_z {\frac{\partial B_z }{\partial y}} \right) - {\frac{\left(\gamma-1\right) \eta }{ {\mu}{\rho_0}}}  \frac{\partial  }{\partial y} \left[ \left(  \frac{\partial B_z }{\partial y} \right) ^2 \right] \label{governing}
\end{eqnarray}

Let us assume that our linear Alfv\'en wave can be represented as:
\begin{eqnarray*}
{{\rm{v}}_z} = \frac{A}{2i} e^{ i \left( \omega t -k y \right)} - \frac{A}{2i} e^{ -i \left( \omega t -k^* y \right)}
\end{eqnarray*}
where $A$ is the amplitude of our wave, $\omega$ is the driving frequency, $k=k_R+ik_I$ is our wavenumber, with complex conjugate $k^*=k_R-ik_I$. Thus, from equation (\ref{v_z}) (with $N_3=0$) we have:
\begin{eqnarray}
b_z = -\frac{\omega B_0 A}{2 {\rm{v}}_A^2 }   \left[ \frac{1}{ik} e^{ i \left( \omega t -k y \right)} - \frac{1}{ik^*} e^{ -i \left( \omega t -k^* y \right)}\right]\label{A0}
\end{eqnarray}

The right-hand-side of equation (\ref{governing}) has two contributions, which can be calculated using our expression for $b_z$. The first term has the form:
\begin{eqnarray}
 {\frac{\partial b_z }{\partial y}} &=& \frac{\omega B_0 A}{2{\rm{v}}_A^2}   \left[ e^{ i \left( \omega t -k y \right)} + e^{ -i \left( \omega t -k^* y \right)}\right]  \nonumber\\
\Rightarrow \frac{\partial  }{\partial t} \left(  b_z {\frac{\partial b_z }{\partial y}} \right) &=&  -\frac{\omega^3 B_0^2 A^2}{2{\rm{v}}_A^4}   \left[ \frac{1}{k} e^{ 2i \left( \omega t -k y \right)} + \frac{1}{k^*} e^{ -2i \left( \omega t -k^* y \right)}\right]\label{A2}
\end{eqnarray}
and the second has the form:
\begin{eqnarray}
&&\left(  \frac{\partial b_z }{\partial y} \right) ^2 = \frac{\omega^2 B_0^2 A^2}{4{\rm{v}}_A^4}   \left[ e^{ 2i \left( \omega t -k y \right)} +  e^{ -2i \left( \omega t -k^* y \right)}  + 2e^{ i \left( k^* - k \right)y}      \right]\nonumber\\
&&\Rightarrow  \frac{\partial  }{\partial y} \left[ \left(  \frac{\partial b_z }{\partial y} \right) ^2 \right] = i \frac{\omega^2 B_0^2 A^2}{2{\rm{v}}_A^4}   \left[ -k e^{ 2i \left( \omega t -k y \right)} + k^*  e^{ -2i \left( \omega t -k^* y \right)} \right.\nonumber\\
&&\left. + \left( k^* -k \right)  e^{ i \left( k^* - k \right)y}      \right]\label{A4}
\end{eqnarray}
The form of equations (\ref{A0}) - (\ref{A4}) lead to equations (\ref{Bz_compare}) - (\ref{hatBF}).

Equations (\ref{A0}) - (\ref{A4}) can also be derived explicitly from our choice of ${\rm{v}}_z$, i.e.:
\begin{eqnarray*}
{{\rm{v}}_z} = A \sin { \left( \omega t -k_R y \right) } e^{ k_I y} 
\end{eqnarray*}
where as before $k=k_R+ik_I$. Thus (from equation \ref{v_z})  $b_z$ has the form:
\begin{eqnarray}
b_z &=& -\frac{\omega B_0 A}{{\rm{v}}_A^2   \left( k_R^2 + k_I^2 \right) }  \left[ k_R \sin { \left( \omega t -k_R y \right) } \right.\nonumber\\
&&\left.-k_I \cos { \left( \omega t -k_R y \right) } \right]  e^{ k_I y} \;.\label{Bzzz}
\end{eqnarray}
Now we can explicitly determine the two contributions:
\begin{eqnarray}
 &&{\frac{\partial b_z }{\partial y}} = \frac{\omega B_0 A}{{\rm{v}}_A^2} \cos { \left( \omega t -k_R y \right) }  e^{ k_I y}  \nonumber\\
&&\Rightarrow \frac{\partial  }{\partial t} \left(  b_z {\frac{\partial b_z }{\partial y}} \right) =  -\frac{\omega^3 B_0^2 A^2}{{\rm{v}}_A^4   \left( k_R^2 + k_I^2 \right) }  \left[ k_R \cos { 2\left( \omega t -k_R y \right) } \right. \nonumber\\
&& \left. + k_I \sin {2 \left( \omega t -k_R y \right) } \right]  {e^{2 k_I y}} \label{A6}
\end{eqnarray}
and the second has the form:
\begin{eqnarray}
&&\left(  \frac{\partial b_z }{\partial y} \right) ^2 = \frac{\omega^2 B_0^2 A^2}{{\rm{v}}_A^4}   \cos^2 { \left( \omega t -k_R y \right) }  e^{ 2 k_I y}\nonumber\\
&&\Rightarrow  \frac{\partial  }{\partial y} \left[ \left(  \frac{\partial b_z }{\partial y} \right) ^2 \right] = \frac{\omega^2 B_0^2 A^2}{{\rm{v}}_A^4}  \left[  k_R \sin { 2\left( \omega t -k_R y \right) } \right.\nonumber \\
&&\left. +k_I \cos {2 \left( \omega t -k_R y \right) } + k_I \right]  e^{ 2 k_I y} \label{A8}
\end{eqnarray}
Equations (\ref{Bzzz}) - (\ref{A8}) give equivalent solutions to  equations (\ref{A0} - \ref{A4}), and thus also lead  to equations (\ref{Bz_compare}) - (\ref{hatBF}).

Note that in both derivations, we have assumed an infinite harmonic solution for ${\rm{v}}_z$, whereas our numerical simulations are for driven harmonic wavetrains. Thus, our solutions are only valid for $c_st < y < {\rm{v}}_A t$.

%----------------------------------------------------------------

\end{appendix}


\begin{thebibliography}{9}
%
\bibitem[2001]{Arber}{{Arber}, T.~D., {Longbottom}, A.~W., {Gerrard}, C.~L., \& {Milne}, A.~M.} 2001, {{J. Comp. Phys.}}, {{171}}, 151
%
\bibitem[2000]{Botha2000}{{{Botha}, G.~J.~J., {Arber}, T.~D., {Nakariakov}, V.~M., \& {Keenan}, F.~P.} 2000, {\aap}, {363}, 1186}
%
\bibitem[1965]{Braginskii1965}{{{Braginskii}, S.~I.} 1965, {{Rev. Plasma Phys.}}, {{1}}, 205}
%
\bibitem[1991]{Browning1991}{{{ {Browning}, P.~K.}} 1991, {Plasma Phys. Cont. Nucl. Fusion}, {33}, 539}
%
\bibitem[1999]{DeMoortel1999}{{{De Moortel}, I., {Hood}, A.~W., {Ireland}, J., \& {Arber}, T.~D.} 1999, {\aap}, {346}, 641}
%
\bibitem[2000]{DeMoortel2000}{{{De Moortel}, I., {Hood}, A.~W., \& {Arber}, T.~D.} 2000, {\aap}, {354}, 334}
%
\bibitem[1983]{Heyvaerts1983}{ { {Heyvaerts}, J., \& {Priest}, E.~R.} 1983, {\aap}, {117}, 220}
%
\bibitem[1971]{Hollweg1971}{ { {Hollweg}, J.~V.} 1971, {\jgr}, {76}, 5155}
%
\bibitem[1988]{Hollweg1988}{ { {Hollweg}, J.~V., \& {Yang}, G.} 1988, {\jgr}, {93}, 5423}
%
\bibitem[2002]{Hood2002}{ { {Hood}, A.~W.,  {Brooks}, S.~J., \& {Wright}, A.~N.} 2002, {Proc. Roy. Soc}, {A458}, 2307}
%
\bibitem[1997a]{Hood1997a}{ { {Hood}, A.~W.,  {Ireland}, J., \& {Priest}, E.~R.} 1997, {\aap}, {318}, 957}
%
\bibitem[1997b]{Hood1997b}{ { {Hood}, A.~W., {Gonzal\'{e}s-Delgado}, D., \& {Ireland}, J.} 1997, {\aap}, {324}, 11}
%
\bibitem[1978]{Ionson1978}{{Ionson, J.~A.} 1978, {\apj}, {226}, 650}
%
\bibitem[1996]{Ireland1996}{{Ireland, J.} 1996, {Ann. Geophys.}, {14}, 485}
%
\bibitem[1986]{Lee1986}{{Lee}, M.~A., \& Roberts, B.} 1986, {{\apj}}, {{301}}, 430
%
\bibitem[1996]{Malara1996}{{{ {Malara}, F., {Primavera}, L., \& {Veltri}, P.}} 1996, {\apj}, {459}, 347}
%
\bibitem[2008]{Mocanu2008}{ {{Mocanu}, G., {Marcu}, A., {Ballai}, I., \& {Orza}, B.} 2008, {Astronomische Nachrichten}, {329}, {780}}
%
\bibitem[1990]{Narain1990}{{{Narain}, U., \& {Ulmschneider}, {P.} 1990, {\ssr}, {54}, 377}}
%
\bibitem[1996]{Narain1996}{{{Narain}, U., \& {Ulmschneider}, {P.} 1996, {\ssr}, {75}, 453}}
%
\bibitem[1995]{Nakariakov1995}{{{{Nakariakov}, V.~M., \& {Oraevsky} V.~N.}} 1995, {\solphys}, {160}, 289}
%
\bibitem[1997]{Nakariakov1997}{{{{Nakariakov}, V.~M., {Roberts} B., \& {Murawski}, K.}} 1997, {\solphys}, {175}, 93}
%
\bibitem[1998]{Nakariakov1998}{{{{Nakariakov}, V.~M., {Roberts} B., \& {Murawski}, K.}} 1998, {\aap}, {332}, 795}
%
\bibitem[1998]{Ofman1998}{{Ofman}, L., {Klimchuk}, J.~A., \& {Davila}, J.~M.} 1998, {\apj}, {{493}}, 474
%
\bibitem[1995]{Ofman1995}{{Ofman}, L., \&  {Davila}, J.~M.} 1995, {\jgr}, {100}, 23413
%
\bibitem[1997]{Ofman1997}{{Ofman}, L., \&  {Davila}, J.~M.} 1997, {\apj}, {476}, 357
%
\bibitem[1991]{Parker1991}{{Parker}, E.~N.} 1991, {\apj}, {376}, 355
%
\bibitem[1996]{Poedts1996}{{Poedts}, S., \& {Boynton}, G.~C.} 1996, {\aap}, {306}, 610
%
\bibitem[1982]{Priest1982}{{Priest}, E.~R.} 1982, {{Solar magnetohydrodynamics}}, {D.~Reidel Publishing.~Company}
%
\bibitem[1998]{Ruderman1998}{{{{Ruderman}, M.~S., {Nakariakov}, V.~M., \& {Roberts} B.}} 1998, {\aap}, {338}, 1118}
%
\bibitem[2002]{RR2002}{{{{Ruderman}, M.~S., \& {Roberts} B.}} 2002, {\apj}, {577}, 475}
%
\bibitem[2007]{Smith2007}{{Smith}, P.~D., {Tsiklauri}, D., \& {Ruderman}, M.~S.} 2007, {\aap}, {475}, 1111
%
\bibitem[2010]{Threlfall2010}{{Threlfall}, J., {McClements}, K.~G., \& {De Moortel}, I.} 2010, {\aap}, {525}, A155 
%
\bibitem[2001]{Tsiklauri2001}{{Tsiklauri}, D., {Arber}, T.~D., \& Nakariakov, V.~M.} 2001, {\aap}, {379}, 1098
%
\bibitem[2002]{TN2002}{{Tsiklauri}, D., \& Nakariakov, V.~M.} 2002, {\aap}, {393}, 321
%
\bibitem[2002]{Tsiklauri2002}{{Tsiklauri}, D., Nakariakov, V.~M., \& {Arber}, T.~D.} 2002, {\aap}, {395}, 285
%
\bibitem[2003]{Tsiklauri2003}{{Tsiklauri}, D., Nakariakov, V.~M., \& {Rowlands}, G.} 2003, {\aap}, {400}, 1051
%
\bibitem[1991]{Ulmschneider1991}{{{Ulmschneider}, P., {Priest}, E.~R., \& {Rosner}, R.} 1991, {{Mechanisms of Chromospheric and Coronal Heating}}, Berlin: Springer}
%
\bibitem[1999]{VerwichteTHESIS1999}{{Verwichte}, E.} 1999, {PhD Thesis: Aspects of Nonlinearity and Dissipation in Magnetohydrodynamics}, {The Open University}
%
\bibitem[1999]{VNL1999}{{Verwichte}, E., {Nakariakov}, V.~M., \& Longbottom, A.~W.} 1999, {J. Plasma Phys.}, {62}, 219
%
\bibitem[1980]{Wilkins1980}{{Wilkins}, M.~L.} 1980, {{J. Comp. Phys.}}, {{36}}, 281
%
\end{thebibliography}
\end{document}